\documentclass[a4paper,11pt]{article}
\pdfoutput=1 

\usepackage{jheppub} 

\usepackage[T1]{fontenc} 
\usepackage{subcaption}
\usepackage{verbatim}
\usepackage{mathrsfs}
\usepackage{physics}
\usepackage{xcolor}
\usepackage{amsthm}
\usepackage{mathtools}
\usepackage{enumerate}

\newcommand{\Del}{\nabla}
\newcommand{\del}{\partial}

\renewcommand{\hat}{\widehat}
\renewcommand{\bar}{\overline}

\newcommand{\comps}{\mathbb{C}}

\newcommand{\A}{\mathcal{A}}
\newcommand{\M}{\mathcal{M}}
\renewcommand{\H}{\mathcal{H}}
\newcommand{\K}{\mathcal{K}}

\newcommand{\D}{\mathcal{D}}
\newcommand{\R}{\mathcal{R}}

\definecolor{indigo(dye)}{rgb}{0.0, 0.25, 0.42}

\newtheorem{theorem}{Theorem}

\newtheorem{lemma}[theorem]{Lemma}

\theoremstyle{definition}

\numberwithin{theorem}{section}

\title{Uniqueness of null-local modular flow}
\author{Gautam Satishchandran}
\author{and Jonathan Sorce}
\affiliation{Princeton Gravity Initiative, Princeton University}
\abstract{
In arXiv:2306.01837, it was conjectured that one can engineer a large class of quantum field theory states for which the modular flow on a spacelike slice is ``instantaneously local.''
Here we show that on null slices, such flows are highly constrained; ultraviolet universality essentially requires null-local modular flow to be unique.
Concretely, we study massless Klein-Gordon theory in Minkowski spacetime, and construct, for any sufficiently regular future-directed vector field on the past null boundary of a causal diamond, a state that has this vector field as its instantaneous modular flow.
We then show by explicit computation that no two distinct states in this class can be realized in the same local Hilbert space.
Using a more abstract argument, we also show that in the vacuum sector of the theory, the \textit{only} null-local modular flow in a causal diamond is provided by the vacuum state itself.
We also comment on the construction of null-local modular flow for massive scalars, free Maxwell fields, free gravitons, and in curved backgrounds.
}
\begin{document}
\maketitle

\section{Introduction}

Many interesting situations in physics involve a restricted subset of a system's total degrees of freedom --- for example the exterior of a black hole, or a collection of qubits that interact with exterior noise.
In discrete systems, the mathematical tools used to describe such restrictions are tensor-factorizing Hilbert spaces and reduced density matrices.
In continuum systems, where tensor factorizations are unavailable, the appropriate tools are those of modular theory.

For a quantum state $|\Psi\rangle$ on a tensor-factorizing space $\H = \H_{\mathcal{R}} \otimes \H_{\mathcal{R}'},$ the modular operator is defined in terms of reduced density matrices as
\begin{equation} \label{eq:modular-factorizing}
\Delta_{\Psi,\mathcal{R}} \equiv \rho_{\mathcal{R}}^{(\Psi)} \otimes \bigl(\rho_{\mathcal{R}^{\prime}}^{(\Psi)}\bigr)^{-1}.
\end{equation}
The ``Tomita-Takesaki theory'' --- see e.g. the reviews in \cite{Witten:notes, Sorce:intuitive} --- allows $\Delta_{\Psi, \mathcal{R}}$ to be defined even when $\R$ cannot be represented as a tensor factor of $\H$.
This theory therefore plays a central role in the analysis of subsystems in continuum quantum theories.
The applications of modular theory are numerous, and we will not attempt to give a complete account of them here. Notable examples include structural results for entropy in quantum field theory \cite{Casini:Bekenstein, Bousso:1,Bousso:2, Cardy:2D-modhams, Hollands:book, Casini:null-plane-modhams, Casini:a-theorem, Lashkari:constraining, Lashkari:sewing, Furuya:mixing, vanLuijk:2024ygl}, proofs of energy conditions \cite{Faulkner:ANEC, Casini:null-plane-modhams, Balakrishnan:QNEC-1, Ceyhan:QNEC-2, Hollands:QNEC-3}, holographic bulk reconstruction \cite{Jafferis:relative, Dong:reconstruction, Faulkner:reconstruction, Faulkner:toolkit, Jefferson:comments, Chen:island, Faulkner:conditional-expectation, Jafferis:observer-1, Levine:seeing, Leutheusser:causal, Gao:observer-2, Leutheusser:times, deBoer:observer-3, Leutheusser:duality, Engelhardt:transfer, Gao:modular, Parrikar:relational, Gesteau:horizons}, and the analysis of generalized entropy in semiclassical gravity \cite{Witten:crossed-product, Chandrasekaran:deSitter, Chandrasekaran:microcanonical, Penington:JT-1, Sorce:types, Kolchmeyer:JT, Jensen:general-subregions, AliAhmad:2023etg, Kudler-Flam:black-holes,Kudler-Flam:approximation, Akers:state-counting, Faulkner:GSL, Kudler-Flam:cosmology, Chen:clock, Kolchmeyer:horizon, Penington:JT-2, Sorce:classical,Klinger:2026tws}.

Despite their utility, explicit expressions for the modular operator are known only in special cases.
The most famous example is due to Bisognano and Wichmann \cite{Bisognano:first,Bisognano:second}, who showed that when $\mathcal{R}$ is a Rindler wedge in Minkowski spacetime, then the modular operator of the vacuum $|\Omega\rangle$ is related to the boost operator $K$:
\begin{equation}
\label{eq:BW-modular}
\Delta_{\Omega,\mathcal{R}} = e^{-2\pi K}.
\end{equation}
Other explicit formulas are known for ``Hartle-Hawking'' states in the presence of static Killing horizons \cite{Sewell:PCT, Wall:GSL,Faulkner:ANEC, Casini:null-plane-modhams}, for conformal rescalings thereof in the case of a general conformal field theory (CFT) \cite{Hislop:1981uh, Casini:CHM, Frob:deSitter}, and also for thermal states \cite{Borchers:thermal, Wong:thermal, Cardy:2D-modhams} or Virasoro descendants of the vacuum \cite{Caminiti:2D} in two-dimensional CFT.
There are also some explicit formulas that are special to the two-dimensional free fermion theory \cite{Casini:fermions, Longo:fermions, Klich:fermions, Hollands:fermions, Blanco:fermions, Fries:fermions, Erdmenger:fermions, Cadamuro:fermions}.

The unifying feature of the above examples is that the ``modular flow'' $\Delta_{\Psi, \mathcal{R}}^{is}$ implements a geometrically local transformation of spacetime.\footnote{In the free fermion case, the transformations may be ``geometrically $n$-local.''}
Intuitively, it makes sense that these are the settings in which it is easiest to find explicit formulas for $\Delta_{\Psi, \mathcal{R}}$ --- if the spacetime symmetries of a theory are sufficiently well understood, then it isn't too big of a leap to understand the states whose modular operators implement those symmetries.
On the other hand, paying \textit{too} much attention to states with geometric modular flow places serious limitations on our ability to apply modular theory to more general subregions or curved backgrounds. It was shown in \cite{Keyl:1993ye, Sorce:analyticity} that any geometrically local modular flow must implement a conformal symmetry of the spacetime background.
Since most spacetimes admit \textit{no} conformal symmetries, most spacetimes cannot support states with geometric modular flow.

\begin{figure}
	\centering
	\includegraphics{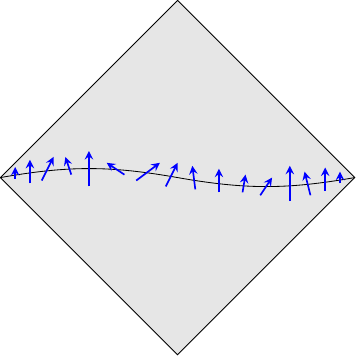}
	\caption{A general domain of dependence with a choice of slice on which we have sketched a vector field that may potentially be the modular flow of a state. The modular flow will not appear local on any other slice of the domain.}
	\label{fig:spacelike-slice-boost}
\end{figure}

Given the utility of modular flow in studying quantum field theory, it seems desirable to find states in generic spacetime backgrounds for which the modular operator admits an explicit, physically useful formula.
However, for the reasons we have just stated, requiring the modular flow to be geometrically local throughout $\mathcal{R}$ is too restrictive.
To get around this problem, the authors of \cite{Jensen:general-subregions} conjectured the generic existence of a large class of states with ``instantaneously local modular flow'' --- i.e., states $|\Psi\rangle$ for which the modular flow is approximately local in the vicinity of a specially chosen Cauchy slice of $\R$, but which is nonlocal on any other slice.
For a sketch of this scenario, see figure \ref{fig:spacelike-slice-boost}.
If such states can be constructed, then modular theory in general backgrounds and for general subregions would acquire a status much closer to that of the special cases discussed above.

\begin{figure}
	\centering
	\includegraphics{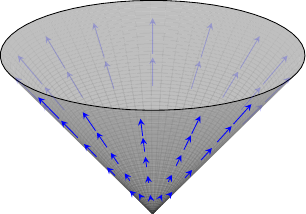}
	\caption{If a region $\R$ admits a null Cauchy slice, then a future-directed vector field on that slice maps the slice into itself. This makes it easier to study instantaneously local modular flows. The slice shown here is the past null boundary $\mathcal{C}^-$ of a Minkowski causal diamond $\mathcal{D}$. The vector field shown here corresponds to the modular flow of the vacuum state in a conformal field theory.}
	\label{fig:cone-flow-vacuum}
\end{figure}

While instantaneously geometric modular flows are not subject to the stringent no-go theorems that preclude a globally geometric modular flow, their general existence is an open question.
For spacelike $\Sigma,$ the key difficulty is that the geometric modular flow of an operator initially localized on $\Sigma$ will immediately leave $\Sigma$; this causes issues both in formulating a precise definition of ``instantaneous locality,'' and in engineering states that might implement this flow.
The situation simplifies if $\mathcal{R}$ is a region that admits a \textit{null} Cauchy slice, as shown in figure \ref{fig:cone-flow-vacuum}; in this case, a future-directed vector field maps the null slice into itself.
In fact, there are already a few known cases of modular flows that are local on a null slice but that do not extend locally to all of $\mathcal{R}$; these include null deformations of a static Killing horizon \cite{Wall:GSL, Faulkner:ANEC, Casini:null-plane-modhams}, as well as the ``Unruh state'' of an out-of-equilibrium black hole \cite{unruh1976notes,Dimock:1987hi,2009arXiv0907.1034D,Gerard:2020tdo,Brum:2014nea,Hollands:2019whz,Klein:2022jtb, Kudler-Flam:black-holes, Kudler-Flam:cosmology}.

The purpose of the present paper is to initiate a more general study of null-local modular flow.
We study the construction of such flows in the simplest possible setting: a free, massless scalar field $\phi$, where the region $\R$ is taken to be a unit-radius causal diamond $\mathcal{D}$ in $D$-dimensional Minkowski spacetime.
The null Cauchy slice in question will be $\mathcal{C}^-,$ which is the past boundary of $\mathcal{D}$.
Because the theory is conformally invariant, the global vacuum state $\Omega$ is known to have geometrically local modular flow in all of $\mathcal{D}$ \cite{Hislop:1981uh, Casini:CHM}; in particular, this means that it has geometrically local modular flow on $\mathcal{C}^-.$
Our main results are:
\begin{enumerate}[(i)]
    \item We construct, for any (sufficiently regular, future-directed, angle-preserving) one-parameter family of diffeomorphisms on $\mathcal{C}^-$, a Gaussian state of the massless scalar field for which the modular flow implements the chosen diffeomorphism.
    These are all legal states in the sense that they provide consistent sets of correlation functions for the field algebra, but (as we are about to explain) they need not all lie in a physical Hilbert space of the theory.
    \item For $D \geq 3$, we show that for any two \textit{distinct} such flows on $\mathcal{C}^-$, the corresponding states can never be excited out of one another using local operators in $\D$.
    In particular, this implies that no non-vacuum state in our family can be realized in the vacuum sector of the theory.
    \item In fact, within the vacuum sector of the theory, the \textit{only} null-local modular flow is that of the vacuum itself, even if we allow the modular flow to be implemented by a weight.
    In other words, result (ii) holds even if Gaussianity, normalizability, and flow-regularity are relaxed.
\end{enumerate}
Results (ii)-(iii) can be thought of as a uniqueness theorem for null-local modular flows.
Intuitively, uniqueness holds because every point on $\mathcal{C}^-$ has zero distance from the edge of $\mathcal{D}$, so changing the modular flow \textit{anywhere} on $\mathcal{C}^-$ induces a violent change in the short-distance singular structure of the state.
Such a change cannot be implemented by the action of local operators.

Before proceeding to the plan of the paper, we aim to give the reader an overview of the key tools we will use in our arguments.
First, because free scalar field theory is obtained by quantizing a linear wave equation, all of the observables in $\mathcal{D}$ can be represented explicitly in terms of initial data on $\mathcal{C}^-$.
For technical reasons, the initial data on $\mathcal{C}^-$ is most easily represented in terms of the ``rescaled descendant fields'' $\del_u \Phi(u, x^{\perp}).$
These are defined via the formula 
\begin{equation}
    \Phi(u, x^{\perp}) = u^{(D-2)/2} \phi(u, x^{\perp}),
\end{equation}
where $u$ is the nontrivial light-cone coordinate on $\mathcal{C}^-,$ and $x^{\perp}$ labels the directions along the transverse sphere.
The global Minkowski vacuum state is determined via Wick contractions, starting with the two-point function 
\begin{equation}
\label{eq:Omega0intro}
\omega_\text{vac}\left(
\partial_{u}\Phi(u_{1},x^{\perp}_{1})\partial_{u}\Phi(u_{2},x^{\perp}_{2})\right) = -\frac{1}{4\pi} \frac{
\delta_{\mathbb{S}^{D-2}}(x_{1}^{\perp},x_{2}^{\perp})}{(u_{1}-u_{2}-i\epsilon)^{2}}.
\end{equation}
With $u$ normalized to lie in the range $[0, 1]$, the modular flow of this state corresponds to a geometric transformation of $\del_u \Phi$ by the integral curves of the vector field $2 \pi u(1-u) (\del / \del u).$

In the sections below, by generalizing equation \eqref{eq:Omega0intro}, we construct, for any appropriately chosen diffeomorphism flow $\psi_s$ on $\mathcal{C}^-$, a two-point function
\begin{equation}
    \omega_{\psi} \left( \del_u \Phi(u_1, x_1^{\perp}) \del_u \Phi(u_2 ,x_2^{\perp}) \right)
        = -\frac{1}{4\pi} \frac{\delta_{\mathbb{S}^{D-2}}(x_{1}^{\perp},x_{2}^{\perp})}{(u_{1}-u_{2}-i\epsilon)^{2}} + S_{\psi}(u_1, x_1^{\perp}; u_2, x_2^{\perp}),
\end{equation}
with $S_{\psi}$ a kernel chosen so that $\omega_{\psi}$ satisfies the ``KMS thermality condition'' with respect to the flow $\psi_s.$
The associated Gaussian state gives rise to a Hilbert space sector $\H_{\omega_\psi}$, and basic theorems from Tomita-Takesaki theory guarantee that the modular flow of the state $|\omega_{\psi}\rangle \in \H_{\psi}$ implements the geometric flow $\psi_s$ on the operators $\del_u \Phi$; this establishes result (i).
For result (ii), we make use of the ``excitability theory'' developed in \cite{Caminiti:2026ewu} based on earlier work in \cite{Powers:quasi, VanDaele:quasi, Araki:quasi, Araki-Yamagami}.
The point is that in order to modify the geometric flow of $\omega_\text{vac}$ to that of $\omega_{\psi},$ the ``correction'' kernel $S_{\psi}$ must be highly singular, and in particular must be too singular to be excited from the vacuum using local operators.\footnote{In fact the same basic idea holds for any two distinct states $\omega_{\psi}$ and $\omega_{\psi'}$, and it was not necessary to assume that one of them was the vacuum.}
Finally, to establish result (iii), we appeal to the theory of the Connes cocycle derivative --- if we \textit{assume} the existence of a weight implementing a null-local modular flow within the vacuum sector $\H_{\text{vac}}$, we can show that $\H_{\text{vac}}$ must also contain a Gaussian state with a null-local modular flow that is different from that of the vacuum (and generically also different from the modular flow of the weight), thereby contradicting result (ii).

The plan of the paper follows.
\begin{itemize}
	\item In section \ref{sec:review}, we review relevant background material: the algebraic quantization of Klein-Gordon theory, the isomorphism between the algebra in the causal diamond $\mathcal{D}$ and the ``initial data'' algebra on $\mathcal{C}^-,$ and some basic definitions and theorems from modular theory.
	\item In section \ref{sec:constructing-states}, we construct, for any one-parameter family of sufficiently-regular, future-directed, angle-preserving diffeomorphisms of $\mathcal{C}^-$, a Gaussian state of massless Klein-Gordon theory for which this diffeomorphism is the modular flow.
	\item In section \ref{sec:no-states}, we show that no two distinct states in the class from section \ref{sec:constructing-states} can be excited out of one another using local operators in the causal diamond.
    We use this result to show that with the exception of the massless vacuum itself, none of these states can be realized in the vacuum sector.
	\item In section \ref{sec:no-weights}, we use the Connes cocycle theorem to show that the vacuum state provides the only null-local flow for the causal diamond in the vacuum sector --- i.e., the results of section \ref{sec:no-states} do not change if one relaxes regularity, Gaussianity, or normalizability.
    \item In section \ref{sec:discussion}, we sketch some generalizations of the ideas introduced in this paper, both to more general field theories and to more general spacetime backgrounds.    
    We comment on the possibility of finding a state with null-local modular flow for the Minkowski diamond within the massive vacuum sector --- which is a nontrivial task because the massive vacuum does not, itself, have null-local modular flow --- and also for finding such states for small causal diamonds in general curved backgrounds. The construction of analogous states for free Maxwell and graviton fields is detailed in appendix \ref{app:nulllocalemgrav}. 
\end{itemize}

\section{Background material}
\label{sec:review}

It is common to think of a quantum system as coming automatically equipped with a Hilbert space.
This perspective, however, obscures the fact that a single quantum field theory admits many different Hilbert space ``sectors'' even on a fixed spacetime background.
Indeed, as explained in the introduction, the purpose of this paper is to construct states with null-local modular flow and to study how their sectors are related. 
To address these questions it is far more useful to begin, instead, with an abstract ``$*$-algebra of fields'' $\A_0,$ then to study the representations of this algebra.

Here we review this perspective for the concrete example of a real Klein-Gordon field, with special attention paid to the ``null quantization'' procedure for a causal diamond in Minkowski spacetime.
We then present a few notions from modular theory that are used elsewhere in the paper.
Our comments on these subjects will be brief.
We refer the reader to \cite{Wald:book,Hollands:review, Caminiti:2026ewu,Witten:notes, Sorce:intuitive} for detailed proofs of all relevant statements. 

\subsection{Algebraic Klein-Gordon theory}
\label{sec:KG-review}

Given a globally hyperbolic spacetime $\M$, one defines an algebra of \textit{smeared fields} that represent the formal objects $\int d^{D} x\, \sqrt{-g}\, \phi(x) f(x).$
Formally, we say that a function $f$ is in the space of \textit{test functions} $C^{\infty}_0(\M)$ if $f$ is a smooth, complex-valued function with compact support.
For each test function $f,$ we define an object $\phi[f]$.
We define a dagger operation by
\begin{equation}
	\phi[f]^{\dagger} = \phi[f^*],
\end{equation}
and we define addition and scalar multiplication by
\begin{equation}
	\phi[f_1] + \phi[f_2] = \phi[f_1+f_2], \qquad \lambda \phi[f] = \phi[\lambda f].
\end{equation}
We then construct the abstract algebra $\A_0$ by taking all formal polynomials in these symbols, and imposing two additional relations that we will now describe.

The first constraint is the canonical commutation relation
\begin{align} \label{eq:CCR}
    \begin{split}
	[\phi[f_1], \phi[f_2]]
        & = - i \int d^{D}x\, d^{D}y\, \sqrt{-g(x)}\, \sqrt{-g(y)}\, f_1(x) E(x, y) f_2(y) \\
        & \equiv - i E[f_1,f_2],
    \end{split}
\end{align}
where $E(x,y)$ is the advanced-minus-retarded propagator for the Klein-Gordon equation.
This is the covariant version of the equal-time commutation relation $[\varphi, \pi] = i \delta.$
In Minkowski spacetime in $D$ dimensions, for example, one has
\begin{equation}
	E(x,y)
	= \frac{i}{2 (2\pi)^{D-1}} \int \frac{d \vec{k}}{\omega_{\vec{k}}} e^{i \vec{k} \cdot (\vec{x} - \vec{y})} \left(e^{- i \omega_{\vec{k}} (x_0 - y_0)} - e^{i \omega_{\vec{k}} (x_0 - y_0)} \right)
\end{equation}
with $\omega_{\vec{k}} = \sqrt{|\vec{k}|^2 + m^2},$ which one can check satisfies the Klein-Gordon equation and reproduces the correct equal-time commutation relations for $\varphi$ and $\pi.$

The second relation imposed on $\A_0$ is the equation of motion.
One wishes to impose $(\Box - m^2)\phi = 0.$
In terms of smeared fields, this is the equation
\begin{equation}
	\int f (\Box - m^2) \phi = 0.
\end{equation}
Integrating by parts, we see that this is formally equivalent to
\begin{equation}
	\int \phi (\Box - m^2) f = 0,
\end{equation}
which can be imposed on the abstract algebra $\A_0$ by setting $\phi[g] = 0$ whenever $g$ can be written as $g = (\Box - m^2) f$ for another test function $f.$

A \textit{sector} of free field theory on the spacetime $\M$ is a Hilbert space representation of the abstract algebra $\A_0.$
A convenient way to produce representations is via the GNS construction.
First, one considers an abstract algebraic state $\omega : \A_0 \to \comps,$ which is a complex-linear functional satisfying the positivity condition
\begin{equation}
	\omega(a^{\dagger} a) \geq 0\, \quad \forall a \in \A_0.
\end{equation}
One then creates an abstract vector space of the formal elements
\begin{equation}
	\text{span}\{|a\omega\rangle\, |\, a \in \A_0\},
\end{equation}
and imposes the inner product
\begin{equation}
	\langle a \omega | b \omega\rangle \equiv \omega(a^{\dagger} b).
\end{equation}
Finally, one takes a quotient by null states and takes the Hilbert space completion, producing a Hilbert space $\H_{\omega}$ on which $\A_0$ is represented.
For details, see e.g. \cite[chapter 1.7]{conway2000course}.
An important class of states for the Klein-Gordon theory are {\em Gaussian states}, for which the connected $n$-point correlation functions vanish for $n\geq 3$.
A Gaussian state is called ``zero-mean'' if it has vanishing one-point function, in which case all of its correlators are determined from the two-point function via Wick contractions.
Note that for any free field state $\omega,$ using the triangle inequality and Cauchy-Schwarz, one automatically has
\begin{equation} \label{eq:positivity-constraint}
    |E[f,g]|
        \leq 2 \sqrt{\omega(\phi[f]^2)} \sqrt{\omega(\phi[g]^2)}, \qquad f, g \text{ real}.
\end{equation}
Conversely, one can actually show (see e.g. \cite[appendix A.7]{Caminiti:2026ewu}) that if $\omega$ is only assumed to be a general Gaussian functional, then inequality \eqref{eq:positivity-constraint} is all that is needed to guarantee positivity, i.e., to guarantee that $\omega$ is a state.

While the fields themselves are represented by unbounded operators on the GNS space $\H_{\omega}$, there is a natural way to complete them into a von Neumann algebra.\footnote{See \cite[appendix A]{Sorce:canonical} for the general construction, or \cite[appendix A]{Caminiti:2026ewu} for the special case of free field theory.}
The resulting von Neumann algebra is called $\A_{\omega}.$
Local subalgebras of $\A_{\omega}$ can be constructed by considering a domain of dependence $\R\subseteq \M$ and constructing the algebra $\A_{\omega}(\R)$ of smeared field operators for which the smearing functions are supported in $\R$.
For any GNS representation, this gives rise to a spacetime net of von Neumann algebras acting on $\H_{\omega}$, which is the basic ingredient in the algebraic approach to quantum field theory \cite{Haag:book}.

\subsection{Initial data formulation}
\label{sec:initial-data-review}

In practice, it is often useful to express smeared field operators in terms of solutions to the Klein-Gordon equation of motion, which can in turn be expressed in terms of initial data.
This is especially important for the present paper, where we need to specify a state in the full causal diamond $\mathcal{D}$ in terms of its behavior on the past null boundary $\mathcal{C}^-.$

The key point is that for every test function $f,$ there is an associated solution to the equation of motion obtained by acting on $f$ with $E$:
\begin{equation}
	Ef(x) \equiv \int d^{D} y\, \sqrt{-g(y)}\, E(x,y) f(y).
\end{equation}
Therefore we can view $E:C_{0}^{\infty}(\M) \to \mathscr{S}$ as a map from the space of test functions to the space $\mathscr{S}$ of smooth solutions with compact support on any spacelike Cauchy slice.
It is known (see e.g. \cite[lemma 3.2.1]{Wald:book}) that this map is surjective, with kernel exactly the
image of $(\Box - m^{2})$.
So each smeared field operator $\phi[f]$ can be thought of as depending only on the solution $Ef$, since this ``change of variables'' automatically implements the relation $\phi[(\Box - m^2) f] = 0$.
Using the fact that $Ef$ can be expressed in terms of initial data on a Cauchy surface $\Sigma$, it follows that the field algebra $\mathcal{A}_0(\mathcal{R})$ in a region $\R$ can be expressed in terms of data defined on any Cauchy surface $\Sigma$ of $\R$.

The explicit relation between $\phi[f]$ and $\Sigma$ is obtained by rewriting the classical expression $\int \phi(x) f(x)$, for $\phi \in \mathscr{S}$, in terms of the initial data for $\phi$ on $\Sigma.$
A simple calculation (see appendix \ref{app:solution-smearing} for details and a figure) gives the classical formula
\begin{equation} \label{eq:general-solution-smearing}
    \phi[f]
        = \int_{\Sigma} d^{D-1} y\, \sqrt{h} (\phi\, n^a \Del_a E f - Ef\, n^a \Del_a \phi),
\end{equation}
where $h$ is the induced metric on $\Sigma$ and $n^a$ is the future-pointing unit normal vector. 
If $\Sigma$ is spacelike, then the initial data for a solution $\chi \in \mathscr{S}$ is specified by ``position data'' $\chi\vert_\Sigma$ and ``momentum data'' $n^a \Del_a \chi|_{\Sigma}$.
Equation \eqref{eq:general-solution-smearing} expresses $\phi[f]$ as a smearing of the position data for $\phi$ against the momentum data for $Ef,$ and of the momentum data for $\phi$ against the position data for $Ef.$

Some regions admit null Cauchy slices.
If we take $\Sigma$ to be such a slice, then the interpretation of equation \eqref{eq:general-solution-smearing} changes; in this case the vector $n^a$ is tangent to $\Sigma,$ so $n^a \Del_a \phi\vert_{\Sigma}$ is determined by $\phi\vert_{\Sigma}.$
In the special case of a Minkowski causal diamond $\D$, we may take $\Sigma$ to be the past null boundary $\mathcal{C}^-.$
Writing $u$ for the null affine parameter scaled to lie in the range $[0,1]$, and writing $x^{\perp}$ for the transverse spherical coordinates, equation \eqref{eq:general-solution-smearing} becomes
\begin{equation}
\label{eq:null-solution-smearing}
    \phi[f] =  \int_{\mathcal{C}^{-}} du\, d\Omega \, u^{D-2}(\phi \partial_{u}Ef - Ef \partial_{u}\phi),
\end{equation}
where $d \Omega = \sqrt{q} d\Omega$ is the volume element associated to the round metric $q_{AB}$ on $\mathbb{S}^{D-2}$.
This expression is more conveniently expressed in terms of the ``rescaled'' fields
\begin{equation}
\label{eq:PhiFflat}
\Phi(u,x^{\perp}) \equiv u^{(D-2)/2}\phi(u,x^{\perp}) \quad \textrm{ and} \quad F(u,x^{\perp}) = u^{(D-2)/2}Ef(u, x^{\perp}).
\end{equation}
Substituting these expressions into equation \eqref{eq:null-solution-smearing}, and integrating by parts in $u,$ we obtain\footnote{To perform the integration by parts, one uses (i) that $Ef$ is continuous at $u=0,$ and (ii) the support of $Ef$ on $\mathcal{C}^-$ is entirely in the past of the support of $f$, so that when $f$ is compactly supported in the interior of $\mathcal{D}$, there is some $0 < u_c < 1$ such that $Ef\vert_{\mathcal{C}^-}$ vanishes for $u \geq u_c.$}
\begin{equation}
\label{eq:phifC1}
	\phi[f]
	= -2 \int_{\mathcal{C}^{-}}  du\, d\Omega\, ~F(u,x^{\perp}) \partial_{u}\Phi(u,x^{\perp}).
\end{equation}

\subsection{Relevant tools from modular theory}
\label{sec:modular-review}

Within a GNS representation $\H_{\omega}$ of an abstract $*$-algebra $\A_0,$ there is a natural procedure (see section \ref{sec:KG-review}) for ``completing'' $\A_0$ into a von Neumann algebra $\A_{\omega}.$
The main advantages of this procedure are that von Neumann algebras (i) are made up of \textit{bounded} operators, and (ii) have useful topological and analytical properties.
In particular, given a von Neumann algebra $\A$ acting on a Hilbert space $\H$, and given a state $|\Psi\rangle \in \H,$ one can define the \textit{modular operator} $\Delta_{\Psi}$ via standard formulas reviewed e.g. in \cite{Witten:notes, Sorce:intuitive}.\footnote{The modular operator is sometimes written $\Delta_{\Psi, \A}$ to emphasize that it depends on the choice of algebra $\A$, but we will leave $\A$ implicit to simplify notation.}
Two key properties of the modular operator are that it preserves the state from which it is constructed,
\begin{equation}
    \Delta_{\Psi} |\Psi\rangle = |\Psi\rangle,
\end{equation}
and that its unitary exponentiation generates a flow on $\A$ called the \textit{modular flow}:
\begin{equation}
    \Delta_{\Psi}^{-is} a \Delta_{\Psi}^{is} \in \A, \qquad a \in \A.
\end{equation}

The physical interpretation of the modular operator comes from the fact that $|\Psi\rangle$ appears thermal within $\A$ when ``time'' is measured with respect to the modular flow parameter.
For a detailed explanation of the intuition behind this statement, see \cite[section 3]{Sorce:intuitive}.
The concrete mathematical statement is that modular flow satisfies the KMS condition, which is that if we fix $a, b \in \A$ and define the function
\begin{equation}
    G_{a b}(i s)
        \equiv \langle \Psi | \Delta_{\Psi}^{-is} a \Delta_{\Psi}^{is} b |\Psi\rangle, \qquad a, b \in \A,
\end{equation}
then this function has an analytic continuation to the complex strip with real values in the range $[0,1],$ such that the boundary value on the other edge of the strip is given by
\begin{equation}
    G_{a b}(1 + i s)
        \equiv \langle \Psi | b \Delta_{\Psi}^{-is} a \Delta_{\Psi}^{is} |\Psi\rangle.
\end{equation}
The \textit{KMS uniqueness theorem}, reviewed in \cite[section 3]{Sorce:intuitive}, states that modular flow is the \textit{only} flow on $\A$ that preserves the state and that satisfies the KMS condition.
This theorem will be used later in the paper to identify the modular flows of certain explicitly constructed states in free field theory.

The last tool we will need is the \textit{Connes cocycle}, which, given two distinct states $|\Psi\rangle$ and $|\Xi\rangle,$ provides a unitary element of $\A$ that ``interpolates'' between the corresponding modular flows.
There are many different conventions for defining the Connes cocycle. The one we choose is that it is a one-parameter family of unitary operators $w_{\Psi | \Xi}(s)$ satisfying
\begin{equation} \label{eq:CC-defining-equation}
    w_{\Psi | \Xi}(s)^{\dagger} \left( \Delta_{\Xi}^{-is} a \Delta_{\Xi}^{is} \right) w_{\Psi | \Xi}(s)
        = \Delta_{\Psi}^{-is} a \Delta_{\Psi}^{is}.
\end{equation}
The construction of the Connes cocycle and of its basic properties (including the fact that it is an element of $\A$) can be found in the textbook account \cite[chapter 8.3]{Takesaki:II}.
Key facts include the cocycle composition law
\begin{equation} \label{eq:cocycle-composition}
    w_{\Psi | \Xi}(s_1 + s_2) = (\Delta_{\Xi}^{- i s_1} w_{\Psi | \Xi}(s_2) \Delta_{\Xi}^{i s_1}) w_{\Psi | \Xi}(s_1),
\end{equation}
together with the fact that $w_{\Psi|\Xi}(s)$ is continuous, as a function of $s,$ with respect to the strong operator topology.

One important feature of modular theory is that modular flows are defined not just for states, but for \textit{weights}.
We will not give explicit definitions until section \ref{sec:no-weights}, but weights are essentially generalized states obtained by relaxing the assumption of normalizability.
For a general (faithful, normal, semifinite) weight $\tau$, there is no modular operator $\Delta_{\tau}$, but there is still a modular flow in the form of a group of automorphisms $\alpha_{s}^{\tau} : \A \to \A$ satisfying an analogue of the KMS condition, and such that $\alpha_{t}^{\tau}$ reduces to conjugation by $\Delta_{\tau}^{-is}$ if $\tau$ happens to be normalizable.
As explained in \cite[chapter 8.3]{Takesaki:II}, a Connes cocycle derivative exists for general (faithful, normal, semifinite) weights, with the defining equation
\begin{equation} \label{eq:cocycle-weight-conversion}
    w_{\tau_1|\tau_2}(s)^{\dagger} \alpha^{\tau_2}_s(a) w_{\tau_1 | \tau_2}(s)
        = \alpha^{\tau_1}_s(a).
\end{equation}

\section{Massless Gaussian states with null-local modular flow}
\label{sec:constructing-states}

In this section we construct an infinite family of states of the massless Klein-Gordon field in a Minkowski causal diamond $\mathcal{D}$, with local modular flow on the past null boundary $\mathcal{C}^{-}$.
As noted in the introduction, a simple example of such a state is the vacuum, for which the modular flow coincides with the action of a conformal boost of $\mathcal{D}$.

In section \ref{sec:masslessvacuum}, we review the example of the vacuum state and explain its representation in terms of initial data on $\mathcal{C}^{-}$.
In section~\ref{sec:gencase}, we generalize this example to construct a large class of Gaussian states with geometric modular flow on $\mathcal{C}^{-}$.  

\subsection{The massless vacuum state}
\label{sec:masslessvacuum}

In Minkowski spacetime, the massless vacuum $\omega_{\text{vac}}$ is a zero-mean Gaussian state of the abstract free field algebra $\A_{0}$. Its two-point function is given by
\begin{equation}
\label{eq:mlessvacbulk}
\omega_{\text{vac}}\left(\phi(x)\phi(x')\right)
=
\frac{\Gamma\left(\frac{D-2}{2}\right)}{4\pi^{D/2}}\,
\frac{1}{
\left(- (t - t' - i \epsilon)^2 + |\vec{x} - \vec{x}'|^2 \right)^{\frac{D-2}{2}}}.
\end{equation}
where the distribution is defined by integrating against test functions $f(x)$ and $g(x')$ before taking the limit $\epsilon \to 0^+.$
Passing to Fourier space gives the equivalent expression
\begin{equation}
    \omega_{\text{vac}}(\phi(x) \phi(x'))
        = \frac{1}{2 (2 \pi)^{D-1}} \int \frac{d \vec{k}}{|\vec{k}|} e^{- i |\vec{k}| (t - t')} e^{i \vec{k} \cdot (\vec{x} - \vec{x}^{\prime})},
\end{equation}
where the distribution is defined by smearing the integrand against test functions before integrating with respect to the spatial momentum $\vec{k}.$
From this form for $\omega_{\text{vac}}$, it is straightforward to verify that $\omega_{\text{vac}}$ is consistent with the canonical commutation relations of equation \eqref{eq:CCR}, and that it satisfies the positivity constraint \eqref{eq:positivity-constraint}; $\omega_{\text{vac}}$ therefore extends to a zero-mean Gaussian state on $\A_0.$
While it is harder to show directly from this form of the two-point function, it is true that $\omega_{\text{vac}}$ satisfies the KMS condition with respect to a certain geometric flow generated by a conformal isometry of the causal diamond $\mathcal{D}.$
In light-cone coordinates $u, v \in [0, 1]$, the relevant flow of the field $\phi$, with flow parameter $s,$ is
\begin{align} \label{eq:massless-modflow}
	\begin{split}
	\phi_s(u, v, x^{\perp})
		& =\left(\frac{e^{2 \pi s}}{((1-u)+e^{2 \pi s} u)((1-v)+e^{2\pi s}v)}\right)^{(D-2)/2} \\
		& \qquad \qquad \times \phi\left(\frac{u e^{2 \pi s}}{(1-u) + e^{2 \pi s} u}, \frac{v e^{2 \pi s}}{(1-v) + e^{2 \pi s} v}, x^{\perp} \right).
	\end{split}
\end{align}
The overall rescaling of $\phi$ comes from the fact that the field $\phi$ has conformal dimension $(D-2)/2.$ 
Since the KMS condition for the two-point function is satisfied by this flow, and since the KMS conditions for all $n$-point functions follow from computing Wick contractions, one may conclude as in \cite{Hislop:1981uh, Casini:CHM} that equation \eqref{eq:massless-modflow} specifies the modular flow of $\omega_{\text{vac}}$ on the algebra of the causal diamond.

Obviously, since $\omega_{\text{vac}}$ has geometric modular flow in all of $\mathcal{D},$ it also has geometric modular flow on the past null boundary $\mathcal{C}^-.$
With a view toward generalizing to other null-local modular flows, we will now study an explicit formula for $\omega_{\text{vac}}$ in terms of initial data on $\mathcal{C}^-$.
One major advantage of doing this is that the KMS structure of $\omega_{\text{vac}}$ is much easier to study from its presentation in terms of null initial data.

In section \ref{sec:review}, we derived equation \eqref{eq:phifC1} relating a ``$\mathcal{D}$ smearing'' of $\phi$ with a ``$\mathcal{C}^-$ smearing'' of $\del_u \Phi$ (with $\Phi = u^{(D-2)/2} \phi$).
By applying this map directly to equation \eqref{eq:mlessvacbulk}, it was shown in \cite{Hollands:thesis} and later
independently in \cite{2009arXiv0907.1034D} that the two-point function obeys the formula
\begin{equation} 
\label{eq:massless-vacuum}
\omega_{\text{vac}}\!\left(\phi[f_{1}]\phi[f_{2}]\right)
=
-\frac{1}{\pi} \lim_{\epsilon \to 0^+}
\int du_{1}\,du_{2}\,d\Omega\,
\frac{F_{1}(u_{1},x^{\perp})\,F_{2}(u_{2},x^{\perp}) }
{(u_{1}-u_{2}-i\epsilon)^{2}}\,
\end{equation}
with $F = u^{(D-2)/2} E f,$ which matches equation \eqref{eq:Omega0intro} from the introduction.
Using the distributional identity 
\begin{equation}
\frac{1}{(u_{1}-u_{2}-i \epsilon)^{2}} - \frac{1}{(u_{2}-u_{1}-i \epsilon)^{2}} = -2\pi i\,\delta'(u_{1}-u_{2}),
\end{equation}
one can easily compute
\begin{equation} 
\omega_{\text{vac}}\left(\phi[f_{1}]\phi[f_{2}]\right)
    - \omega_{\text{vac}}\left(\phi[f_2] \phi[f_1] \right)
= - 2 i \int du\,d\Omega\, \left(\del_{u} F_{1}(u,x^{\perp})\right)\,F_{2}(u,x^{\perp}),
\end{equation}
which gives us the initial-data form of the commutator kernel $E$ as
\begin{equation} \label{eq:Cminus-CCR}
    E[f_1, f_2]
        = 2 \int du\, d\Omega \left(\del_{u} F_1(u, x^{\perp})\right) F_2(u, x^{\perp}).
\end{equation}

From equation \eqref{eq:massless-vacuum}, we will now verify the KMS property of $\omega_{\text{vac}}$ directly.
A straightforward calculation shows that the flow \eqref{eq:massless-modflow} acts, on $\mathcal{C}^-$, as the local diffeomorphism of $\del_u \Phi$ generated by the vector field $\xi = 2 \pi u (1-u) (\del / \del u).$
To study the behavior of $\omega_{\text{vac}}$ under this flow, it is helpful to rewrite equation \eqref{eq:massless-vacuum} in terms of a new coordinate $\eta$ such that we have $\xi = \del / \del \eta.$
The explicit coordinate transformation is
\begin{equation}
    \eta = \frac{1}{2 \pi} \log\frac{u}{1-u},
\end{equation}
and the range of $\eta$ is $- \infty < \eta < \infty.$
Applying this coordinate transformation to $\omega_{\text{vac}}$ and absorbing a positive function into the parameter $\epsilon$, we obtain 
\begin{equation}
\label{eq:2pteta}
\omega_{\text{vac}}\!\left(\phi[f_{1}]\phi[f_{2}]\right)
=
-\pi
\lim_{\epsilon\to 0^{+}}
\int d\eta_{1}\,d\eta_{2}\,d\Omega\,
\frac{
F_{1}(\eta_{1},x^{\perp})
F_{2}(\eta_{2},x^{\perp})
}
{
\sinh^{2}\!\left(\pi(\eta_{1}-\eta_{2}-i\epsilon)\right)
}\, .
\end{equation}
The transformation in equation \eqref{eq:massless-modflow} acts on $\del_u \Phi$ via the translation $\eta \mapsto \eta + s$; since equation \eqref{eq:2pteta} was obtained by smearing the two-point function of $\del_u \Phi$ against $F_1$ and $F_2$ on $\mathcal{C}^-,$ one deduces the flow formula
\begin{equation}
\omega_{\text{vac}}\!\left(\phi_{s_1}[f_{1}]\phi_{s_2}[f_{2}]\right)
=
-\pi
\lim_{\epsilon\to 0^{+}}
\int d\eta_{1}\,d\eta_{2}\,d\Omega\,
\frac{
F_{1}(\eta_{1},x^{\perp})
F_{2}(\eta_{2},x^{\perp})
}
{
\sinh^{2}\!\left(\pi(\eta_{1} + s_1 -\eta_{2} - s_2 -i\epsilon)\right)
}\, .
\end{equation}
From this expression, invariance of the two-point function under simultaneous modular flow of all operators ($s_1 = s_2$) is manifest.

To see the KMS property, we must study the function
\begin{equation} \label{eq:G-position-space}
    G_{f_1, f_2}(i s)
        \equiv \omega_{\text{vac}}(\phi_{s}[f_1] \phi[f_2])
        = - \pi \lim_{\epsilon\to 0^{+}} \int d\eta_{1}\,d\eta_{2}\,d\Omega\,
            \frac{F_{1}(\eta_{1},x^{\perp}) F_{2}(\eta_{2},x^{\perp})}{\sinh^{2}\!\left(\pi(s + \eta_{1} - \eta_{2} -i\epsilon)\right)}.
\end{equation}
This can be analytically continued in $s$ so long as no singularities are introduced in the denominator of the integrand; i.e., to any complex value of $s$ such that we never have $s + \eta_1 - \eta_2 - i \epsilon$ equal to $i$ times an integer.
This allows us to analytically continue $s$ into the strip with imaginary part in the range $(-1, 0]$.
Computing the boundary value $G_{f_1, f_2}(1 + i s)$ is subtle, because to actually compute $G_{f_1,f_2}(x + i s)$ using equation \eqref{eq:G-position-space}, one must take $\epsilon$ smaller than $1-x$ to avoid singularities. To compute the boundary value, it is convenient to transform equation \eqref{eq:2pteta} to momentum space using the formula
\begin{equation} \label{eq:sinh-Fourier}
	\frac{1}{\sinh^2(\pi (\eta_1 - \eta_2) - i \epsilon)}
	= - \frac{1}{2 \pi^2} \int_{-\infty}^{\infty} d\nu\, \frac{\nu e^{- \frac{\nu}{2} + \frac{\epsilon \nu}{\pi}}}{\sinh(\nu/2)} e^{i \nu(\eta_1 - \eta_2)}.
\end{equation}
Plugging this into the formula for $\omega_{\text{vac}}$ gives
\begin{align}
\label{eq:omega0nueta}
	\begin{split}
		\omega_{\text{vac}}(\phi[f_1] \phi[f_2])
		= \frac{1}{2 \pi} \int d\nu\, d\eta_{1}\,d\eta_{2}\,d\Omega~ &  \frac{\nu e^{-\nu/2}}{\sinh(\nu/2)} e^{i \nu (\eta_1 - \eta_2)} F_{1}(\eta_{1},x^{\perp})F_{2}(\eta_{2},x^{\perp}).
	\end{split}
\end{align}
In this representation, we have
\begin{equation} \label{eq:G-momentum-space}
    G_{f_1, f_2}(x + i s)
        = \frac{1}{2 \pi} \int d\nu\, d\eta_{1}\,d\eta_{2}\,d\Omega~
        \frac{\nu e^{-\nu/2}}{\sinh(\nu/2)} e^{i \nu (\eta_1 - \eta_2 + s)} e^{x \nu} F_{1}(\eta_{1},x^{\perp})F_{2}(\eta_{2},x^{\perp}).
\end{equation}
Plugging in $x=1$ and substituting $\nu \to -\nu$ gives the desired KMS property
\begin{equation}
    G_{f_1, f_2}(1 + i s)
        = \omega_{\text{vac}}(\phi[f_2] \phi_s[f_1]).
\end{equation}

Finally, we note that we can obtain a particularly useful formula from 
equation \eqref{eq:omega0nueta} by rewriting it in terms of the Fourier transformed fields 
\begin{equation}
\label{eq:F0hat}
\hat{F}_{\text{vac}}(\nu,x^{\perp}) = \int_{-\infty}^{\infty}d\eta~e^{-i\nu  \eta}F(\eta,x^{\perp}),
\end{equation}
where the subscript ``vac'' is there to remind us that the coordinate $\eta$ is adapted to the modular flow of $\omega_{\text{vac.}}$.
Writing \eqref{eq:omega0nueta} in terms of $\hat{F}_{\text{vac}}$, we  obtain 
\begin{align} \label{eq:omega-0-frequency}
		\omega_{\text{vac}}(\phi[f_{1}] \phi[f_{2}])
		= \frac{1}{2 \pi} \int d\nu d\Omega~ \frac{\nu e^{- \nu/2}}{\sinh(\nu/2)} \hat{F}_{\text{vac},1}(-\nu, x^{\perp}) \hat{F}_{\text{vac},2}(\nu, x^{\perp}).
\end{align}
So we see that $\omega_{\text{vac}}$ has the form of an $L^2$ inner product for the functions $\hat{F}_{\text{vac},1}$ and $\hat{F}_{\text{vac},2}$, modified by a diagonal kernel in frequency space.

\subsection{Gaussian states with general null-local modular flow}
\label{sec:gencase}

We now construct an infinite family of Gaussian states with geometric modular flow on $\mathcal{C}^{-}$.
To begin, let $\psi_s$ be a one-parameter family of angle-preserving diffeomorphisms of $\mathcal{C}^-\cong (0,1)\times \mathbb{S}^{d-2}$ into itself, acting as $u\mapsto \psi_{s}(u,x^{\perp})$ for all $s\in \mathbb{R}$.
We assume that the generator of $\psi_s$ is future directed and non-vanishing on $(0,1)\times \mathbb{S}^{d-2}$.
The flow on fields will be defined on $\mathcal{C}^-$ by $\Phi_s(u, x^{\perp}) = \Phi(\psi_s(u, x^{\perp}), x^{\perp}),$ with $\Phi = u^{(D-2)/2} \phi.$
Subject to a regularity condition on $\psi_s$ that we will define shortly (see equation \eqref{eq:regcond}), we will construct a Gaussian state $\omega_{\psi}$ on $\mathcal{C}^-$ that satisfies the KMS condition with respect to $\psi_s.$
Once we give an explicit formula for the correlation functions of $\omega_{\psi}$, the key points we must check are (i) compatibility with the commutation relations; (ii) positivity of $\omega_{\psi}$ as a functional of the fields; and (iii) the KMS property.

In the spirit of the preceding subsection, we begin by writing the infinitesimal generator of $\psi_s$ as a vector field
\begin{equation}
\label{eq:zeta}
    \zeta^a = \chi(u, x^{\perp}) \left( \frac{\del}{\del u}\right)^a.
\end{equation}
We then define a coordinate $\rho$ such that we have $\zeta = \del / \del \rho.$
Concretely, we may choose
\begin{equation}
\label{eq:rho}
\rho = \int_{1/2}^{u} \frac{du^{\prime}}{\chi(u^{\prime},x^{\perp})}.
\end{equation}
Equivalently, $\rho$ is the coordinate that solves the formula
\begin{equation} \label{eq:rho-as-diff}
    \psi_{\rho}(1/2, x^{\perp}) = (u, x^{\perp}),
\end{equation}
from which it is clear that $\rho$ has range $(-\infty, \infty)$.
Inspired by equation \eqref{eq:2pteta} for the massless vacuum, we define a two-point function $\omega_{\psi}$ via the formula
\begin{equation} \label{eq:omegapsi-ansatz}
    \omega_{\psi}(\phi[f_1] \phi[f_2])
        = - \pi \lim_{\epsilon \to 0^+} \int d\rho_1\,d\rho_2\, d\Omega \frac{F_1(\rho_1, x^{\perp}) F_2(\rho_2, x^{\perp})}{\sinh^2(\pi(\rho_1 - \rho_2 - i \epsilon))},
\end{equation}
with $F = u^{(D-2)/2} Ef.$
We produce a candidate state on $\mathcal{C}^-$ by defining all higher-point functions by Wick contraction.
To show that this is actually a well defined state, we need to show that equation \eqref{eq:omegapsi-ansatz} is finite for all choices of $f_1$ and $f_2$, that it respects the canonical commutation relations, and that it satisfies the positivity condition.

For several of these points, it is helpful to work in momentum space, as in equation \eqref{eq:omega0nueta}.
Writing
\begin{equation}
    \hat{F}_{\psi}(\nu, x^{\perp})
        = \int_{-\infty}^{\infty} d\rho\, e^{-i\nu \rho} F(\rho, x^{\perp}),
\end{equation}
equation \eqref{eq:omegapsi-ansatz} is equivalent to
\begin{align} \label{eq:omega-psi-frequency}
		\omega_{\psi}(\phi[f_{1}] \phi[f_{2}])
		= \frac{1}{2 \pi} \int d\nu\, d\Omega~ \frac{\nu e^{- \nu/2}}{\sinh(\nu/2)} \hat{F}_{\psi,1}(-\nu, x^{\perp}) \hat{F}_{\psi,2}(\nu, x^{\perp}).
\end{align}
From this expression, we will now determine a regularity condition on $\psi_s$ to guarantee finiteness.

The frequency kernel $\nu e^{-\nu/2} / \sinh(\nu/2)$ is smooth at $\nu=0,$ exponentially decaying in the limit $\nu \to \infty,$ and grows linearly in $\nu$ in the limit $\nu \to -\infty.$
The function $F(\rho, x^{\perp})$ is automatically smooth in $\rho,$ so its Fourier transform decays faster than any rational function.
Consequently, finiteness of equation \eqref{eq:omega-psi-frequency} at infinity does not require any special conditions on $\psi$.
Finiteness at $\nu=0$, on the other hand, requires that each $\hat{F}_{\psi}$ be square-integrable, i.e., that the original functions $F(\rho, x^{\perp}) = u^{(D-2)/2} Ef$ be square-integrable.
The support properties of $Ef$ are such that $F$ vanishes for sufficiently large $\rho,$ meaning that square integrability of $F$ is determined entirely by its behavior in the limit $\rho \to -\infty.$
The function $Ef$ is finite in this limit,\footnote{Note that in even dimensions, propagation of massless fields happens purely along the light cone, and $Ef(\rho, x^{\perp})$ will actually vanish for $\rho$ less than some critical value. In this case one does not need to impose any additional regularity condition on $\psi_s$.} so the only question is whether $u^{(D-2)/2}$ is square integrable.
To guarantee finiteness of \eqref{eq:omega-psi-frequency}, we therefore impose the condition
\begin{equation}
\label{eq:regcond}
\textrm{ $u^{\frac{D-2}{2}}(\rho,x^{\perp})$ is square-integrable with respect to $d\rho\, d\Omega$ at $\rho =-\infty$.}
\end{equation}
This condition is extremely mild; for example, in the vacuum case when we have $u(\rho) = e^{2 \pi \rho}/(1 + e^{2 \pi \rho}),$ the regularity condition is satisfied due to the exponential decay of $u$ in the limit $\rho \to -\infty.$

Well definedness of our formula for $\omega_{\psi}$ is now established under the regularity condition \eqref{eq:regcond}.
Next, we must show that $\omega_{\psi}$ respects the canonical commutation relations.
For this, we employ equation \eqref{eq:omega-psi-frequency} to compute the formula
\begin{equation} \label{eq:omega-psi-commutator}
    \omega_{\psi}(\phi[f_1] \phi[f_2] - \phi[f_2] \phi[f_1])
        = -\frac{1}{\pi} \int d\nu\, d\Omega~ \nu \hat{F}_{\psi,1}(-\nu, x^{\perp}) \hat{F}_{\psi,2}(\nu, x^{\perp}).
\end{equation}
On the other hand, from equation \eqref{eq:Cminus-CCR}, we have
\begin{equation}
    E[f_1, f_2] = 2 \int du\, d\Omega \left(\del_{u} F_1(u, x^{\perp})\right) F_2(u, x^{\perp}),
\end{equation}
and we may change to the $\rho$ coordinate as
\begin{equation}
    E[f_1, f_2] = 2 \int d\rho\, d\Omega \left(\del_{\rho} F_1(\rho, x^{\perp})\right) F_2(\rho, x^{\perp}).
\end{equation}
Clearly, after Fourier transforming and comparing to equation \eqref{eq:omega-psi-commutator}, we get the desired relation
\begin{equation}
    \omega_{\psi}(\phi[f_1] \phi[f_2] - \phi[f_2] \phi[f_1])
        = - i E[f_1, f_2].
\end{equation}

Our next step is to check that $\omega_{\psi}$ actually defines a \textit{state} on the algebra of fields in the diamond.
This is the requirement that $\omega_{\psi} : \A_0(\D) \to \comps$ is positive.
As reviewed in section \ref{sec:review}, positivity of $\omega_{\psi}$ is equivalent to an inequality relating $\omega_{\psi}$ to the commutator; one must have
\begin{equation}
	|E[f, g]| \leq 2 \sqrt{\omega_{\psi}(\phi[f]^2)} \sqrt{\omega_{\psi}(\phi[g]^2)}
\end{equation}
for generic real test functions $f$ and $g.$
But from our above calculations, we have
\begin{equation}
	|E[f,g]|
		= \frac{1}{\pi} \left| \int d\nu\, d\Omega\, \nu\, \hat{F}_{\psi}(-\nu, \Omega) \hat{G}_{\psi}(\nu, \Omega) \right|,
\end{equation}
and the Cauchy-Schwarz inequality gives
\begin{equation}
|E[f,g]|
	\leq \frac{1}{\pi} \sqrt{\int d\nu\, d\Omega\, |\nu|\, |\hat{F}_{\psi}(\nu, \Omega)|^2} \sqrt{\int d\nu\, d\Omega\, |\nu|\, |\hat{G}_{\psi}(\nu, \Omega)|^2},
\end{equation}
hence 
\begin{equation}
|E[f,g]|
	\leq 2 \sqrt{\frac{1}{2 \pi} \int d\nu\, d\Omega\, |\nu|\, |\hat{F}_{\psi}(\nu, \Omega)|^2} \sqrt{\frac{1}{2\pi} \int d\nu\, d\Omega\, |\nu|\, |\hat{G}_{\psi}(\nu, \Omega)|^2}.
\end{equation}
Comparing this to the expression
\begin{equation}
	\omega_{\psi}(\phi[f]^2)
		= \frac{1}{2 \pi} \int_{\nu, \Omega} \frac{\nu e^{- \nu/2}}{\sinh(\nu/2)} |\hat{F}_\psi(\nu, \Omega)|^2,
\end{equation}
where we may freely symmetrize over $\nu \to -\nu$ in the integrands to obtain
\begin{equation}
	\omega_{\psi}(\phi[f]^2)
	= \frac{1}{2 \pi} \int_{\nu, \Omega} |\nu| |\coth(\nu/2)| |\hat{F}_\psi(\nu, \Omega)|^2,
\end{equation}
one finds that inequality \eqref{eq:positivity-constraint} follows from the general inequality $|\coth(\nu/2)| \geq 1.$

Finally, we must verify that $\omega_\psi$ satisfies the KMS condition with respect to the flow $\Phi_s(u, x^{\perp}) = \Phi_s(\psi_s(u,x^{\perp}), x^{\perp}).$
By similar arguments as in the previous subsection, one easily finds the position- and frequency-space formulas
\begin{equation}
    \omega_{\psi}(\phi_{s_1}[f_1] \phi_{s_2}[f_2])
        = - \pi \lim_{\epsilon \to 0^+} \int d\rho_1\,d\rho_2\, d\Omega \frac{F_1(\rho_1, x^{\perp}) F_2(\rho_2, x^{\perp})}{\sinh^2(\pi(\rho_1 + s_1 - \rho_2 - s_2 - i \epsilon))}
\end{equation}
and
\begin{align}
		\omega_{\psi}(\phi[f_{1}] \phi[f_{2}])
		= \frac{1}{2 \pi} \int d\nu\, d\eta_1\, d\eta_2\, d\Omega~ \frac{\nu e^{- \nu/2}}{\sinh(\nu/2)} e^{i \nu (\eta_1 - \eta_2)} F_{1}(\rho_1, x^{\perp}) F_2(\rho_2, x^{\perp}).
\end{align}
From the position-space expression, one finds invariance under simultaneous modular flow of all arguments ($s_1 = s_2$), and sees that the expression
\begin{equation}
    G_{f_1, f_2}(i s)
        = \omega_{\psi}(\phi_{s}[f_1] \phi[f_2])
\end{equation}
may be analytically continued to $x + is$ for $0 \leq x < 1.$
From the momentum-space expression, one may easily verify the boundary value
\begin{equation}
    G_{f_1, f_2}(1 + i s)
        = \omega_{\psi}(\phi[f_2] \phi_{s}[f_1]),
\end{equation}
completing the proof of the KMS condition.
The KMS uniqueness theorem reviewed in section \ref{sec:modular-review} therefore shows that the flow generated by $\psi_{s}$ is the modular flow of the causal diamond algebra with respect to the state $\omega_{\psi}$. 

\section{Uniqueness of null-local modular flow for Gaussian states}
\label{sec:no-states}

In the preceding section, we constructed a large family of Gaussian states $\omega_{\psi}$ for which the modular flow reproduces an angle-preserving one-parameter family of diffeomorphisms $u \mapsto \psi_s(u, x^{\perp})$.
In this section we show that {\em none} of these states can lie in the vacuum sector $\mathcal{H}_{\text{vac}}$.
To prove this, we first review in section \ref{subsec:revexcitability} the theory of ``excitability'' developed in \cite{Caminiti:2026ewu}.
Applying this theory to the problem at hand, we show in section \ref{sec:violating} that for $\psi_s \neq \psi'_s,$ the states $\omega_{\psi}$ and $\omega_{\psi'}$ cannot be excited from one another by local operators.
In section \ref{sec:vacuum-uniqueness}, we explain how this general result implies that none of the states in $\H_{\text{vac}}$ can reproduce the correlation functions of $\omega_{\psi}$ within the causal diamond $\mathcal{D}$, except in the trivial case $\omega_{\psi} = \omega_{\text{vac}}.$

Before proceeding, a few comments are in order about the framework.
An algebraic state is defined with respect to a particular algebra; the states we have been calling $\omega_{\psi}$ and $\omega_{\text{vac}}$ are defined on the causal diamond algebra $\A_0(\mathcal{D}).$
It happens to be the case that $\omega_{\text{vac}}$ has a natural extension to the Poincar\'{e}-invariant vacuum on the full algebra $\A_0(\M)$ defined over all of Minkowski spacetime.
The ``vacuum sector'' $\H_{\text{vac}}$ is the space obtained by performing the GNS construction for the vacuum state defined over all of $\A_0(\M)$; this is an \textit{extension} of the state we have been calling $\omega_{\text{vac}}$ so far.
We will write $\H_{\omega_{\text{vac}}}$ for the GNS space defined from the vacuum using only operators in $\mathcal{D}$; this matches the notation from section \ref{sec:KG-review}.
The space $\H_{\omega_{\text{vac}}}$ is actually dense within the full vacuum sector $\H_{\text{vac}}$ thanks to the Reeh-Schlieder property of the vacuum \cite{Reeh:1961ujh}, and this density will play an important role in section \ref{sec:vacuum-uniqueness}; in reading sections \ref{subsec:revexcitability} and \ref{sec:violating}, however, one should keep the restriction to $\A_0(\mathcal{D})$ in mind.

\subsection{Review of excitability for Gaussian states}
\label{subsec:revexcitability}

Each of the states $\omega_{\psi}$ constructed in section \ref{sec:gencase} is defined on the $*$-algebra $\A_0(\mathcal{D})$, and gives rise to a GNS space $\H_{\omega_{\psi}}$ via the formal equation
\begin{equation}
    \H_{\omega_{\psi}}
        = \bar{\A_0(\mathcal{D}) |\omega_{\psi} \rangle}.
\end{equation}
Following terminology introduced in \cite{Caminiti:2026ewu}, we say that $\omega_{\psi'}$ can be \textit{excited} out of $\omega_{\psi},$ written $\omega_{\psi'} \prec \omega_{\psi},$ if there is a vector or density matrix in $\H_{\omega_{\psi}}$ that reproduces the correlation functions of $\omega_{\psi'}$ within $\mathcal{D}$.
Since the physical role of the GNS space $\H_{\omega_{\psi}}$ is to organize the space of local excitations around $\omega_{\psi}$, the mathematical relation $\omega_{\psi'} \prec \omega_{\psi}$ precisely captures the physical concept of excitability.
We will now review the theorem of \cite{Caminiti:2026ewu} characterizing excitability for zero-mean Gaussian states; see also \cite{Caminiti:excitability-2} for the case of nonzero mean, and \cite{Powers:quasi, VanDaele:quasi, Araki:quasi, Araki-Yamagami} for related earlier work.

For the excitability relation $\omega_{\psi'} \prec \omega_{\psi}$ to hold, three conditions must be checked.
Since both $\omega_{\psi}$ and $\omega_{\psi'}$ are zero-mean Gaussian, these conditions can be expressed entirely in terms of the two-point functions of the states.
The starting point is the ``symmetrized two-point function'' $\mu_{\psi}$, defined on the space of test functions as
\begin{equation}
    \langle f | g \rangle_{\mu_{\psi}}
        \equiv \frac{\omega_{\psi}(\phi[f^*] \phi[g] + \phi[g] \phi[f^*])}{2}.
\end{equation}
This is a semi-inner product on the space of test functions.
The Hilbert space obtained by taking the quotient by null states, then performing a completion, is called $\K_{\mu_{\psi}}.$

The first necessary condition for excitability ($\omega_{\psi'} \prec \omega_{\psi}$) is that the $\mu_{\psi'}$ inner product must be bounded in terms of the $\mu_{\psi}$ inner product, meaning that there must exist a constant $C$ with 
\begin{equation}
\label{eq:domcond}
\bra{f}\ket{f}_{\mu_{\psi'}} \leq C \bra{f}\ket{f}_{\mu_{\psi}}, \qquad f \text{ real}.
\end{equation}
This implies that the $\mu_{\psi'}$ inner product can be represented on $\mathcal{K}_{\mu_{\psi}}$ by a bounded operator $Q_{\psi|\psi'}$, defined via the formula
\begin{equation}
\label{eq:Qpsi}
\bra{f_{1}}Q_{\psi|\psi'}\ket{f_{2}}_{\mu_{\psi}} \equiv \bra{f_{1}}\ket{f_{2}}_{\mu_{\psi'}}.
\end{equation}
On the other hand, even without imposing condition \eqref{eq:domcond}, the positivity constraint \eqref{eq:positivity-constraint} implies the existence of an operator $R_{\psi}$ on $\K_{\mu_{\psi}}$ representing the commutator:
\begin{equation}
    \langle f_1 | R_{\psi} | f_2 \rangle_{\mu_{\psi}} \equiv \frac{1}{2} E[f_1^*, f_2].
\end{equation}
One can view the positive, self-adjoint operators $Q_{\psi|\psi^{\prime}}-i R_{\psi}$ and $1-iR_{\psi}$ as the ``operator representations'' of $\omega_{\psi}$ and $\omega_{\psi^{\prime}}$ on $\mathcal{K}_{\mu_{\psi}}$, as matrix elements of these operators reproduce the corresponding two-point functions.

The remaining two conditions for excitability are related to properties of the operators $Q_{\psi|\psi'}$ and $R_{\psi}.$
Once condition \eqref{eq:domcond} is satisfied, excitability $\omega_{\psi'} \prec \omega_{\psi}$ is equivalent to the pair of conditions
\begin{enumerate}[(i)]
    \item \label{cond1} $Q_{\psi|\psi'}$ has trivial kernel.
    \item \label{cond2}The difference of the operators $\sqrt{Q_{\psi|\psi'} - i R_{\psi}}$ and $\sqrt{1 - i R_{\psi}}$ is Hilbert-Schmidt on $\mathcal{K}_{\mu_{\psi}}$:
    \begin{equation} \label{eq:HS-condition}
        \tr\left[\left( \sqrt{Q_{\psi|\psi'} - i R_{\psi}} - \sqrt{1 - i R_{\psi}} \right)^2 \right] < \infty.
    \end{equation}
\end{enumerate}

As explained in \cite{Caminiti:2026ewu}, condition (i) ensures that one can formulate a candidate density-matrix representative for $\omega_{\psi'}$ on $\mathcal H_{\omega_\psi}$, while condition (ii) guarantees that this candidate is normalizable.
In the following, we will prove $\omega_{\psi^{\prime}}\not \prec \omega_{\psi}$ by showing that condition (ii) is always violated when $\psi_{s}$ and $\psi_{s}^{\prime}$ are distinct.

\subsection{Gaussian states with null-local modular flow are not mutually excitable}
\label{sec:violating}

Now we show, using the tools from the preceding subsection, that one always has $\omega_{\psi'} \not\prec \omega_{\psi}$ for $\psi^{\prime}_s \neq \psi_s.$
We note that to disprove excitability, it will be sufficient to consider the restriction of $\mu_{\psi}$ to the subspace of real test functions.
So, for simplicity, all test functions will be real throughout the remainder of this subsection.

Our strategy of proof will be to first obtain explicit formulas for $\mu_{\psi'}$ and $\mu_{\psi}$, and then to show that they must violate the finite-trace condition \eqref{eq:HS-condition}.
Our starting point is the frequency-space representation \eqref{eq:omega-psi-frequency}, which, after symmetrization, gives the formula
\begin{equation}
    \langle f_1 | f_2 \rangle_{\mu_{\psi}}
        = \frac{1}{2\pi} \int d\nu\, d\Omega\, \nu \coth(\nu/2)
        \bar{\hat{F}_{\psi,1}(\nu, x^{\perp})} \hat{F}_{\psi,2}(\nu, x^{\perp}).
\end{equation}
Here we are using the notation
\begin{equation}
    F = u^{(D-2)/2} Ef,
    \qquad
    \hat{F}_{\psi}(\nu, x^{\perp}) = \int_{-\infty}^{\infty} d\rho\, e^{-i \nu \rho} F(\rho, x^{\perp}),
\end{equation}
where recall that $\rho$, defined by equation \eqref{eq:rho-as-diff}, is the coordinate adapted to the flow $\psi_{s}$.
We will rewrite this suggestively as
\begin{equation} \label{eq:mu-psi}
    \langle f_1 | f_2 \rangle_{\mu_{\psi}}
        = \frac{1}{2\pi} \int d\nu\, d\Omega\,
        \left( \bar{\sqrt{\nu \coth(\nu/2)} \hat{F}_{\psi,1}(\nu, x^{\perp})}\right)
        \left( \sqrt{\nu \coth(\nu/2)} \hat{F}_{\psi,2}(\nu, x^{\perp})\right),
\end{equation}
which makes it clear that $\mu_{\psi}$ is really just an $L^2$ inner product for a certain class of functions, i.e., those functions obtained by Fourier transforming with respect to $\rho$ then multiplying by $\sqrt{\nu \coth(\nu/2)}$.

Using equation \eqref{eq:mu-psi}, we may relate $\mu_{\psi'}$ to $\mu_{\psi}$ by finding a transform between the functions $\hat{F}_{\psi'}$ and $\hat{F}_{\psi}$.
To do this, one must invert the Fourier transform used to define $\hat{F}_{\psi'},$ then re-perform the Fourier transform used to define $\hat{F}_{\psi}.$
The relevant expression is
\begin{align}
    \begin{split}
    \hat{F}_{\psi'}(\nu, x^{\perp})
        & = \int_{-\infty}^{\infty} d\rho'\, e^{- i \nu \rho'} F(\rho', x^{\perp}) \\
        & = \int_{-\infty}^{\infty} d\rho'\, e^{- i \nu \rho'} \int_{-\infty}^{\infty} \frac{d\lambda}{2 \pi} e^{i \lambda \rho} \hat{F}_{\psi}(\lambda, x^{\perp}) \\
        & = \int_{-\infty}^{\infty} d\lambda\, \left( \frac{1}{2\pi} \int_{-\infty}^{\infty} d\rho'\, e^{- i (\nu \rho' - \lambda \rho)}\right) \hat{F}_{\psi}(\lambda, x^{\perp}).
    \end{split}
\end{align}
The term in parentheses may be thought of as defining a kernel $A_{\psi'|\psi}(\nu,\lambda; x^{\perp})$; this is not just the identity kernel, since $\rho'$ is a nontrivial function of $\rho$ --- the coordinate $\rho'$ parametrizes the flow $\psi'_s,$ while the coordinate $\rho$ is parametrizes the flow $\psi_s.$
We write the final expression as
\begin{align}
    \begin{split}
    \hat{F}_{\psi'}(\nu, x^{\perp})
        & = \int_{-\infty}^{\infty} d\lambda\, A_{\psi'|\psi}(\nu, \lambda; x^{\perp}) \hat{F}_{\psi}(\lambda, x^{\perp}).
    \end{split}
\end{align}
From this expression, we obtain a formula for $\mu_{\psi'}$ as
\begin{align}
    \begin{split}
    \langle f_1 | f_2 \rangle_{\mu_{\psi'}}
        = \frac{1}{2\pi} \int d\nu\, & d\Omega\,
        \nu \coth(\nu/2) \\
        & \times \bar{\int d \lambda_1\, A_{\psi'|\psi}(\nu, \lambda_1; x^{\perp}) \hat{F}_{\psi,1}(\lambda_1, x^{\perp})} \\
        & \times \int d \lambda_2\, A_{\psi'|\psi}(\nu, \lambda_2; x^{\perp}) \hat{F}_{\psi,2}(\lambda_2, x^{\perp})
    \end{split}
\end{align}
This can be rearranged as
\begin{align} \label{eq:mu-psiprime}
    \begin{split}
    \langle f_1 | f_2 \rangle_{\mu_{\psi'}}
        = \frac{1}{2\pi} & \int d\lambda_{1}\, d\Omega\,
        \bar{\sqrt{\lambda_1 \coth(\lambda_1/2)} \hat{F}_{\psi,1}(\lambda_1, x^{\perp})} \\
        & \times \sqrt{\lambda_1 \coth(\lambda_1/2)} \int d\nu\, d\lambda_2\, \frac{\nu \coth(\nu/2)}{\lambda_1 \coth(\lambda_1/2)}  \bar{A_{\psi'|\psi}(\nu, \lambda_1; x^{\perp})} A_{\psi'|\psi}(\nu, \lambda_2;x^{\perp}) \\
        & \times \hat{F}_{\psi, 2}(\lambda_2, x^{\perp})
    \end{split}
\end{align}
The operator $Q_{\psi|\psi'},$ if it exists, is supposed to satisfy the equation $\langle f_1 | Q_{\psi|\psi'}|f_2\rangle_{\mu_{\psi}} = \langle f_1 | f_2\rangle_{\mu_{\psi'}}.$
So by comparing equations \eqref{eq:mu-psi} and \eqref{eq:mu-psiprime}, we see that if $Q_{\psi|\psi'}$ exists, then it acts on the function $\hat{F}_{\psi}$ via the kernel
\begin{equation} \label{eq:Q-kernel}
    K(\lambda_1, \lambda_2; x^{\perp})
        = \int d\nu\,\frac{\nu \coth(\nu/2)}{\lambda_1 \coth(\lambda_1/2)} \bar{A_{\psi'|\psi}(\nu, \lambda_1; x^{\perp})} A_{\psi'|\psi}(\nu, \lambda_2;x^{\perp})
\end{equation}

To determine whether the kernel of equation \eqref{eq:Q-kernel} gives rise to a bounded operator on $\K_{\mu_{\psi}},$ one would need to check the domination condition \eqref{eq:domcond}.
However, to prove the excitability obstruction $\omega_{\psi'} \not\prec \omega_{\psi}$, we do not actually need to check boundedness; we will simply show that even \textit{if} the domination condition is satisfied, the finite-trace condition \eqref{eq:HS-condition} is violated, meaning that excitability cannot hold.

The key point is that condition \eqref{eq:HS-condition} is a Hilbert-Schmidt condition.
An operator $L$ is said to be Hilbert-Schmidt if we have $\tr(L^{\dagger} L) < \infty.$
Condition \eqref{eq:HS-condition} is the Hilbert-Schmidt property for the operator $A - B$, with
\begin{align}
    A
        & = \sqrt{Q_{\psi|\psi'} - i R_{\psi}}, \\
    B
        & = \sqrt{1 - i R_{\psi}}.
\end{align}
The Hilbert-Schmidt property is preserved under left or right multiplication by bounded operators.
So if $A - B$ is Hilbert-Schmidt, then so is
\begin{equation}
    \frac{1}{2} \left[ (A-B) (A+B) + (A+B)(A-B) \right]
        = A^2 - B^2.
\end{equation}
So if $A - B$ is Hilbert-Schmidt, then so is $A^2 - B^2 = Q_{\psi|\psi'} - 1.$

To prove $\omega_{\psi'} \not\prec \omega_{\psi},$ it therefore suffices to show that $Q_{\psi|\psi'} - 1$ can never be Hilbert-Schmidt for $\psi'_s \neq \psi_s.$
This follows because the kernel in equation \eqref{eq:Q-kernel} is diagonal in the angular directions; it has the form of a kernel in $\lambda_1$ and $\lambda_2$, multiplied by the delta distribution for the angular directions.
It follows that $Q_{\psi|\psi'} - 1,$ as a kernel, is proportional to the delta distribution on the transverse sphere; we may write this heuristically as 
\begin{equation}
    (Q_{\psi|\psi'} - 1)
        \sim \Lambda(\lambda_1, \lambda_2) \delta_{\mathbb{S}^{D-2}}(x_1^{\perp}, x_2^{\perp}).
\end{equation}
The delta distribution is just the identity operator on an infinite-dimensional function space; it has infinite trace, as do all of its powers.
So the operator $Q_{\psi|\psi'}-1$ can only be Hilbert-Schmidt if it vanishes identically.
But if $Q_{\psi|\psi'} - 1$ vanishes identically, then we have $\mu_{\psi'} = \mu_{\psi}$ and so the symmetric parts of $\omega_{\psi}$ and $\omega_{\psi'}$ match one another.
Since the antisymmetric parts are determined by the canonical commutation relations, vanishing of $Q_{\psi|\psi'} - 1$ forces $\omega_{\psi'} = \omega_{\psi}$.
We conclude, as desired, that we have $\omega_{\psi'} \not\prec \omega_{\psi}$ except in the degenerate case $\omega_{\psi'} = \omega_{\psi}.$

 We collect the results of this section in the following lemma. 
\begin{lemma}
\label{lem:nogaussian}
Let $D\geq 3$, and let $\omega_{\psi}$ and $\omega_{\psi'}$ be Gaussian states on $\mathcal{A}_{0}(\mathcal{D})$ with null-local modular flow in the family constructed in section \ref{sec:gencase}.. Then $\omega_{\psi'}\prec\omega_{\psi}$ if and only if $\omega_{\psi'}=\omega_{\psi}$. 
\end{lemma}

\subsection{Uniqueness of null-local flow for Gaussian states in the vacuum sector}
\label{sec:vacuum-uniqueness}

Thus far, we have shown that for $\omega_{\psi'} \neq \omega_{\psi}$ within the class of Gaussian states constructed in section \ref{sec:gencase}, we have $\omega_{\psi'} \not\prec \omega_{\psi}.$
In particular, this means that for any non-vacuum state $\omega_{\psi}$ in this class, we have $\omega_{\psi} \not\prec\omega_{\text{vac}}.$
However, as emphasized in the preamble to this section, this result is expressed purely in terms of the causal diamond $\mathcal{D}$; there is a subtle distinction between the GNS space $\H_{\omega_{\text{vac}}}$ and the global vacuum sector $\H_{\text{vac}}.$
Here we explain how lemma \ref{lem:nogaussian} implies a stronger claim:
within the global vacuum sector $\H_{\text{vac}},$ there is no density matrix with modular flow given by $\psi_s$ on the past null boundary $\mathcal{C}^-$ of the causal diamond.

Intuitively, lemma \ref{lem:nogaussian} states that $\omega_{\psi}$ cannot be excited out of $\omega_{\text{vac}}$ using only operators localized within $\mathcal{D}$.
The only potential loophole that could arise within the full vacuum sector would be if $\omega_{\psi}$ could be excited out of $\omega_{\text{vac}}$ using operators outside of $\mathcal{D}.$
This is be ruled out by the Reeh-Schlieder property of the vacuum \cite{Reeh:1961ujh}, which states that every state in $\H_{\text{vac}}$ can be approximated arbitrarily well by acting only within $\mathcal{D}.$

For a formal argument, we write $\Omega$ for the global vacuum state on all of Minkowski spacetime.
The vacuum sector $\H_{\text{vac}}$ is the GNS space for $\Omega.$
There is an isometric embedding of $\H_{\omega_{\text{vac}}}$ into $\H_{\text{vac}}$ via the formula
\begin{equation}
    V \left( a |\omega_{\text{vac}}\rangle \right)
        \equiv a |\Omega\rangle, \qquad a \in \A_0(\mathcal{D}).
\end{equation}
The Reeh-Schlieder property implies that $V$ has dense range; so it is in fact a unitary map from $\H_{\omega_{\text{vac}}}$ to $\H_{\text{vac}}$.
Now let us suppose, toward contradiction, that there is a density matrix $\sigma_{\psi}$ on $\H_{\text{vac}}$ that reproduces the correlation functions of $\omega_{\psi}$ within the causal diamond $\mathcal{D}.$
Then $V^{\dagger} \sigma_{\psi} V$ will be a density matrix within $\H_{\omega_{\text{vac}}}.$
One clearly has
\begin{equation}
    \tr_{\H_{\omega_{\text{vac}}}}((V^{\dagger} \sigma_{\psi} V) a)
        = \tr_{\H_{\text{vac}}}(\sigma_{\psi} a), \qquad a \in \A_0(\mathcal{D}).
\end{equation}
So if there is a density matrix in $\H_{\text{vac}}$ reproducing $\omega_{\psi}$ within $\mathcal{D},$ then the Reeh-Schlieder property lets us construct a corresponding density matrix in $\H_{\omega_{\text{vac}}}$, which contradicts lemma \ref{lem:nogaussian}.
We may therefore conclude that there is no state in $\H_{\text{vac}}$ restricting to $\omega_{\psi}$ on the causal diamond, except in the degenerate case $\omega_{\psi} = \omega_{\text{vac}}.$

\section{Uniqueness of null-local modular flow: general states and weights}
\label{sec:no-weights}

In the preceding section we proved that, in the family of Gaussian states with null-local modular flow constructed in section \ref{sec:gencase}, the only state that lies in vacuum sector $\mathcal{H}_{\text{vac}}$ is the vacuum state itself.
However, we have not yet ruled out the existence of states with null-local modular flow that lie outside of this class. 
Moreover, modular flow is not defined just for states; the broadest category of objects for which modular flow can be defined are suitably regular ``weights.''
So we are led to ask: does there exist any weight in the vacuum sector with null-local modular flow, other than the vacuum state itself?\footnote{Every state is a weight, so this question also addresses the possibility of having null-local modular flow for a state outside the class constructed in section \ref{sec:gencase}.}

In this section we prove that the answer is no: any weight in the vacuum sector with modular flow generated by an angle-preserving diffeomorphism $\psi_{s}$ on $\mathcal{C}^{-}$ must be proportional to the vacuum state, and hence be equal to $\omega_{\text{vac}}$ after appropriate normalization.
Moreover, this result holds for any $\psi_{s}$ that generates an automorphism of the algebra; in particular, we do not need to impose the regularity condition \eqref{eq:regcond}.
In the remainder of this section we review the properties of weights, then provide a proof of our claim.
The main results are summarized in theorem \ref{thm:uniquenesstheorem}.

A weight on a von Neumann algebra $\mathcal{A}$ assigns, to each positive element of the algebra, a value in the range $[0, \infty]$.
Formally, the set of positive elements of $\A$ is called $\A_+,$ and a weight $\tau$ is a map $\tau : \A_+ \to [0, \infty]$ that respects addition (i.e., $\tau(a + b) = \tau(a) + \tau(b)$) and multiplication by positive scalars (i.e., $\tau(\lambda a) = \lambda \tau(a)$ for $\lambda\in\mathbb{R}_{+}$).
Any state on $\A$ defines a weight, but weights are more general: a weight can be infinite for some elements of $\A_+.$

The relevant class of weights that we will be interested in are ``faithful, normal, semifinite (FNS)'' weights.
Formal definitions can be found in \cite[chapter 7]{Takesaki:II}.
Conceptually, a weight is called {\em faithful} if $\tau(a)=0$ implies $a=0$; \textit{normality} is a continuity condition appropriate for functions that can take values in $[0, \infty]$; and \textit{semifiniteness} guarantees that the finite-weight elements of $\A_+$ are appropriately dense within $\A_{+}$.
In the context of modular theory --- see for example the textbook account \cite[chapter 8]{Takesaki:II} --- FNS weights play a similar role to states.
In particular, every FNS weight $\tau$ induces a modular flow $\alpha^{\tau}_s : \A \to \A.$

One useful example to keep in mind is that of a \textit{thermal weight} in the Minkowski vacuum sector of a quantum field theory; the thermal ``density matrix'' $e^{-\beta H}$ is non-normalizable due to the continuous energy spectrum, but it still defines a weight via the map
\begin{equation}
    \tau_{\beta}(a)
        = \tr(e^{-\beta H} a).
\end{equation}
The modular flow induced by this weight is just global time translation, with a parametrization that depends on the inverse temperature $\beta.$

Our proof of the excitability lemma \ref{lem:nogaussian}, and the associated claim about uniqueness of null-local modular flow in $\H_{\text{vac}},$ relied crucially on violating a certain normalizability criterion (equation \eqref{eq:HS-condition}).
One may therefore wonder if, the uniqueness of null-local flows in $\H_{\text{vac}}$ only holds within the class of \textit{states}. In other words, in a similar manner to the thermal weight, relaxing to general FNS weights might in principle allow us to realize a non-vacuum flow $\psi_s$ using modular theory.
We will now prove that this is not the case.

To rule this out we assume, toward contradiction, that the vacuum sector admits a faithful, semifinite, normal weight $\tau_{\psi}$ for which the modular flow $\alpha^{\tau_{\psi}}_s$ acts on $\mathcal{C}^-$ according to
\begin{equation}
    \alpha^{\tau_{\psi}}_s(\Phi(u, x^{\perp}))
        = \Phi(\psi_s(u, x^{\perp}), x^{\perp}).
\end{equation}
As in section \ref{sec:initial-data-review}, we have defined $\Phi = u^{(D-2)/2} \phi.$
The flow $\psi_s$ is angle-preserving, future-directed, and has nowhere-vanishing generator; however, we do not impose the regularity condition \eqref{eq:regcond}.
The Connes cocycle theorem, reviewed in section \ref{sec:modular-review}, implies the existence of unitary operators $w_{\tau_{\psi}|\omega_{\text{vac}}}(s)$ in the von Neumann algebra of $\mathcal{D}$, satisfying the equation
\begin{equation} \label{eq:cocycle-interpolation-vac}
    w_{\tau_{\psi}|\omega_{\text{vac}}}(s)^{\dagger}
        \alpha^{\text{vac}}_s(a) w_{\tau_{\psi}|\omega_{\text{vac}}}(s)
        = \alpha^{\tau_{\psi}}_s(a).
\end{equation}
Using this cocycle, we can define a particular family of states within the vacuum sector via the formula
\begin{equation}
    |\Omega_s\rangle
        \equiv w_{\tau_{\psi}|\omega_{\text{vac}}}(s) |\Omega\rangle,
\end{equation}
where $|\Omega\rangle$ is the \textit{global} vacuum state on all of Minkowski spacetime.
To reach a contradiction, we will show that the states $|\Omega_s\rangle$ will induce, when restricted to the diamond, a member of the family constructed in section \ref{sec:gencase}. We then apply lemma \ref{lem:nogaussian} of section \ref{sec:vacuum-uniqueness} to rule out the possibility of $\tau_{\psi}$ being nontrivial.

To understand the state $|\Omega_s\rangle$, we begin by studying how the cocycle acts on the algebra.
By substituting $a \mapsto \alpha_{-s}^{\text{vac}}(\Phi)$ in equation \eqref{eq:cocycle-interpolation-vac}, we find
\begin{equation}
    w_{\tau_{\psi}|\omega_{\text{vac}}}(s)^{\dagger}
        \Phi(u, x^{\perp}) w_{\tau_{\psi}|\omega_{\text{vac}}}(s)
        = \alpha^{\tau_{\psi}}_s(\alpha^{\text{vac}}_{-s}(\Phi(u, x^{\perp}))
        = \Phi(\theta_s(u, x^{\perp}), x^{\perp}),
\end{equation}
where $\theta_s$ is the angle-preserving diffeomorphism obtained by acting first with the vacuum flow for parameter $-s,$ then with the $\tau_{\psi}$ flow for parameter $+s.$
From this, we may write the two-point function of $|\Omega_s\rangle$ on $\mathcal{C}^-$ as
\begin{align}
    \begin{split}
    & \langle \Omega_s | (\del_{u} \Phi)(u_1, x^{\perp}) (\del_{u}\Phi)(u_2, x^{\perp}) |\Omega_s\rangle \\
        & \qquad = \del_{u_1} \theta_{s}(u_1, x^{\perp})
            \del_{u_2} \theta_s(u_2, x^{\perp})
            \times \langle \Omega | (\del_{u} \Phi)(\theta_s(u_1, x^{\perp}), x^{\perp}) (\del_{u} \Phi)(\theta_s(u_2, x^{\perp}), x^{\perp}) |\Omega \rangle.
\end{split}
\end{align}
Using equations \eqref{eq:phifC1} and \eqref{eq:massless-vacuum} for the vacuum two-point function (see also equation \eqref{eq:Omega0intro}), we obtain the formula
\begin{align} \label{eq:Omega-s-2pt}
    \begin{split}
    \langle \Omega_s | \phi[f_1] \phi[f_2] |\Omega_s\rangle
        = - \frac{1}{\pi} \lim_{\epsilon \to 0^+} \int du_1\, du_2\, d\Omega\, & \del_{u_1} \theta_s(u_1,x^{\perp}) \del_{u_2} \theta_s(u_2, x^{\perp}) \\
        & \times \frac{F_1(u_1, x^{\perp}) F_2(u_2, x^{\perp})}{(\theta_s(u_1, x^{\perp}) - \theta_s(u_2, x^{\perp}) - i \epsilon)^2}.
    \end{split}
\end{align}
Moreover, since the cocycle acts geometrically on fields, it is easy to see that all higher-point functions of $|\Omega_s\rangle$ will be determined by Wick contractions; i.e., it is a Gaussian state.

Now we aim to show, from equation \eqref{eq:Omega-s-2pt}, that the restriction of $|\Omega_s\rangle$ to the causal diamond $\mathcal{D}$ is in the class of states constructed in section \ref{sec:gencase}.
Our analysis will be informed by the ansatz that the modular flow of this state should be given by the diffeomorphism $\chi_{s'} = \theta_s^{-1} \circ \varphi_{s'} \circ \theta_s,$ where $\varphi_{s'}$ is the diffeomorphism corresponding to the vacuum modular flow, i.e., the diffeomorphism generated by the vector field $2 \pi u (1-u) (\del/\del u).$
This assumption inspires us to define, as in section \ref{sec:gencase}, a new coordinate $\rho$ via the formula
\begin{equation}
    u
        = (\theta_s^{-1} \circ \varphi_{\rho} \circ \theta_s)(1/2, x^{\perp}).
\end{equation}
Differentiating and using the chain rule gives
\begin{align}
    \begin{split}
    d u
        & = (\del_u \theta_s^{-1})\left((\varphi_{\rho} \circ \theta_s)(1/2, x^{\perp}), x^{\perp}\right) \times (\del_{\rho} \varphi_{\rho})(\theta_s(1/2, x^{\perp}), x^{\perp}) \times d\rho \\
        & = \frac{1}{{(\del_u \theta_s)(u, x^{\perp})}} \times 2 \pi \left((\varphi_{\rho} \circ \theta_s)(1/2, x^{\perp})\right) \left(1 - (\varphi_{\rho} \circ \theta_s)(1/2, x^{\perp})\right) \times d\rho \\
        & = 2 \pi \frac{\theta_s(u, x^{\perp}) (1 - \theta_s(u, x^{\perp}))}{{(\del_u \theta_s)(u, x^{\perp})}} d \rho.
    \end{split}
\end{align}
Substituting $\rho$ for $u$ in equation \eqref{eq:Omega-s-2pt}, and absorbing a positive function into the parameter $\epsilon,$ we obtain the formula
\begin{align}
    \begin{split}
    \langle \Omega_s | \phi[f_1] \phi[f_2] |\Omega_s\rangle
        = - \pi \lim_{\epsilon \to 0^+} \int d\rho_1\, d\rho_2\, d\Omega
        \frac{F_1(\rho_1, x^{\perp}) F_2(\rho_2, x^{\perp})}{\sinh^2(\pi(\rho_1 - \rho_2 - i \epsilon))}.
    \end{split}
\end{align}
So we see explicitly that $|\Omega_s\rangle$ is in the class of states constructed in section \ref{sec:gencase}.
Since it is in the vacuum sector, the results of section \ref{sec:vacuum-uniqueness} give that $|\Omega_s\rangle$, restricted to the causal diamond $\mathcal{D}$, agrees with $|\Omega\rangle.$
In other words, for every operator $a$ localized in $\mathcal{D},$ we have
\begin{equation}
    \langle \Omega | a | \Omega \rangle
        = \langle \Omega_s | a | \Omega_s \rangle
        = \langle \Omega | w_{\tau_{\psi}|\omega_{\text{vac}}}(s)^{\dagger} a w_{\tau_{\psi}|\omega_{\text{vac}}}(s) |\Omega \rangle.
\end{equation}
From this expression, one sees that the cocycle operator commutes, within vacuum correlators, with any operator localized in the diamond.
A classic result known as the \textit{centralizer theorem} --- see e.g. \cite[appendix B.3]{Sorce:canonical} for a simple proof --- then implies that the cocycle operator must be fixed by vacuum modular flow. Since the vacuum modular flow is geometric in the diamond, it is ``ergodic'' in the sense that the only operator invariant under the flow are scalar multiples of the identity. 
This gives
\begin{equation} \label{eq:cocycle-phase}
    w_{\tau_{\psi}|\omega_{\text{vac}}}(s) = \zeta(s), \qquad \zeta(s) \in U(1).
\end{equation}

From equation \eqref{eq:cocycle-phase}, we will now show our desired claim, which is that $\tau_{\psi}$ is proportional to the vacuum state $\omega_{\text{vac}}.$
The point is that the form of $\zeta(s)$ is strongly constrained by the properties of the cocycle that we reviewed in section \ref{sec:modular-review}.
In particular, the composition law of equation \eqref{eq:cocycle-composition} gives
\begin{equation}
    \zeta(s_1 + s_2) = \zeta(s_1) \zeta(s_2).
\end{equation}
This property, together with continuity of $\zeta$, implies the existence of a real number $c$ satisfying
\begin{equation}
    \zeta(s) = e^{i c s}.
\end{equation}
To get from here to a proportionality equation relating $\tau_{\psi}$ and $\omega_{\text{vac}},$ we make use of a KMS property for the cocycle given in \cite[theorem 8.3.3c]{Takesaki:II}.
That theorem (with a few notational substitutions to match our conventions from section \ref{sec:modular-review}) implies the existence, for each operator $a$ in the causal diamond, of an analytic function in the unit-width complex strip with
\begin{align}
    G_a(is)
        & = e^{- i c s} \tau_{\psi}(a) \\
    G_a(1 + i s)
        & = e^{- i c s} \omega_{\text{vac}}(a).
\end{align}
The function $e^{- c (z-1)} \omega_{\text{vac}}(a)$ is analytic in the same domain, and matches the boundary value $G_a(1+is)$; it must therefore also match the boundary value $G_a(is),$ in particular giving
\begin{equation}
    e^{c} \omega_{\text{vac}}(a)
        = G_a(0) = \tau_{\psi}(a).
\end{equation}
We therefore obtain
\begin{equation}
    \tau_{\psi} = e^{c} \omega_{\text{vac}},
\end{equation}
as desired. We summarize the results of this section in the following theorem. 
 \begin{theorem}
\label{thm:uniquenesstheorem}
Let $D\geq 3$ and let $\tau_{\psi}$ be any faithful, normal, semifinite weight on the von Neumann algebra of the causal diamond within $\H_{\text{vac}}$, with the property that the modular flow of $\tau_{\psi}$ on $\mathcal{C}^{-}$ corresponds to an angle-preserving diffeomorphism $\psi_{s}$. Then $\tau_{\psi}=e^{c}\omega_{\text{vac}}$ for some $c\in \mathbb{R}$.
\end{theorem}

\section{Future directions}
\label{sec:discussion}

Our main result has been to show that for the massless Klein-Gordon field in Minkowski spacetime, the vacuum state $\omega_{\text{vac}}$ is the unique state (or weight) with geometric modular flow on the past null boundary $\mathcal{C}^{-}$ of a causal diamond $\mathcal{D}$.
The question remains open, however, as to whether \textit{some} state with null-local modular flow can be found in more general settings.
It would be appealing to be able to construct a state with null-local modular flow for more general QFTs and more general spacetimes, since --- as emphasized in the introduction --- states with geometric formulas for their modular flow have been very useful in studying both quantum field theory and semiclassical quantum gravity.

One obvious generalization to consider is that of a non-conformally invariant field theory.
For example, we could study a free massive scalar, or massless gauge fields like the free graviton or the free Maxwell field (which is non-conformal in $D \neq 4$).
Another important generalization is to consider quantum field theories on generic, curved backgrounds.

In the simplest case of the massive scalar field in Minkowski spacetime, the vacuum state does not have geometric modular flow on $\mathcal{C}^-,$ so even before the question of uniqueness can be broached, one should ask whether there exists \textit{any} state in the vacuum Hilbert space with geometric modular flow on $\mathcal{C}^-.$
We provide some evidence below that the answer is yes, at least in sufficiently low dimensions.
As for curved backgrounds, sufficiently small causal diamonds have past boundaries $\mathcal{C}^-$ with the same basic geometry as in Minkowski spacetime, and one can follow ideas introduced in \cite{Hollands:thesis} to construct correlators that give rise to null-local modular flow; the question is whether these correlators can be realized in physical (i.e. Hadamard) Hilbert spaces.
We sketch this construction below, and also comment on the case of free gauge fields such as the free Maxwell and graviton fields.\footnote{An additional subtlety arises in the gravitational case, where in order to define $\mathcal{A}_{\textrm{GR}}(\mathcal{D})$, the subregion $\mathcal{D}$ must be defined in a diffeomorphism-invariant way. We comment on this in appendix \ref{app:nulllocalgrav}.}

\subsection{Massive scalar field}
\label{subsec:massive}

The simplest nontrivial generalization of our results is to a Klein-Gordon scalar with nonvanishing mass $m.$
The massive vacuum state $\omega_{m}$, restricted to a causal diamond $\mathcal{D},$ can be restricted to $\mathcal{C}^-$ using the techniques of section \ref{sec:initial-data-review}.
The resulting formula is of the schematic form
\begin{equation}
\label{eq:2ptlambdam}
\omega_{m}(\phi[f_{1}]\phi[f_{2}]) = 4\int du_{1}\, d\Omega_{1}\, du_{2}\,d\Omega_{2}~F_{m,1}(u_{1},x_{1}^{\perp})F_{m,2}(u_{2},x_{2}^{\perp})~\lambda_{m}(u_{1},x_{1}^{\perp};u_{2},x_{2}^{\perp}).
\end{equation}
where we define
\begin{equation}
    F_m \equiv u^{(D-2)/2} E_m f,    
\end{equation}
and $E_m$ is the advanced-minus-retarded propagator for the massive scalar.
The kernel $\lambda_m$ takes the form of a mass-dependent correction to the massless answer:
\begin{equation}
\label{eq:lambdam}
\lambda_{m}(u_{1},x_{1}^{\perp};u_{2},x_{2}^{\perp}) = -\frac{1}{4\pi} \frac{\delta_{\mathbb{S}^{D-2}}(x_{1}^{\perp},x_{2}^{\perp})}{(u_{1}-u_{2}-i0^{+})^{2}} + S_{m}(u_{1},x_{1}^{\perp};u_{2},x_{2}^{\perp}).
\end{equation}
The correction term $S_m$ is a symmetric bidistribution on $\mathcal{C}^-$ with weaker singularities than the first term in the above equation. This formula was derived in $D=4$ by Janssen and Verch \cite{Janssen:characteristic}; using the explicit Bessel-function form for the two-point function of the massive vacuum, it is straightforward to generalize their formula to any dimension $D\geq 3$ as in equation \eqref{eq:lambdam} (see also \eqref{eq:Sm-shortdistance} below).

If we were to set the correction term $S_m$ to zero by fiat, we would obtain the two-point function of the massless vacuum; we will call this $\omega_0$ in the present section, though note that it the same as what we have been calling $\omega_{\text{vac}}$ elsewhere in the paper.
The state $\omega_0$ provides a perfectly good two-point function for a Gaussian state in the massive Klein-Gordon theory, \textit{and} it has null-local modular flow just as in section \ref{sec:masslessvacuum}; the essential question is whether it can be realized in the vacuum sector of the massive theory.
In the language of section \ref{sec:no-states}, the question to ask is whether we have the excitability relation $\omega_0 \prec \omega_m$.
It is worth noting that for any of the other states $\omega_{\psi}$ constructed in section \ref{sec:gencase}, the same exact argument as in section \ref{sec:no-states} shows that we have $\omega_{\psi} \not\prec \omega_m$ --- so the only \textit{possible} null-local modular flow in the massive vacuum sector is that of $\omega_0.$

Using the results of \cite{Caminiti:2026ewu}, the excitability relation $\omega_0 \prec \omega_m$ is controlled by details of the relationship between the two-point functions.
These details, in turn, are dependent on the structure of $S_m.$
For excitability to hold, it seems that one will need the singularities of $S_m$ to be sufficiently weak.

Repeating the analysis of \cite{Verch:Hadamard} for general $D\geq 3$ dimensions, it is straightforward to show that $S_{m}$, away from the tip of $\mathcal{C}^-$, is smooth in $u_{1}$ and $u_{2}$ and is singular only at angular coincidence. The rotational invariance of $\omega_{0}$ and $\omega_{m}$ on $\mathcal{C}^{-}$ implies that $S_{m}$ can only depend on the geodesic distance $s_{12}(x_{1}^{\perp},x_{2}^{\perp})$  between the points $x_{1}^{\perp}$ and $x_{2}^{\perp}$ on $\mathbb{S}^{D-2}$. In the limit $s_{12}\to 0$, $S_{m}$ behaves as\footnote{Following \cite{Verch:Hadamard}, the leading angular singularity is equivalent to singular behavior of the Greens function of the Laplacian on $\mathbb{S}^{D-2}$, from which one can derive this formula.} 
\begin{equation}
\label{eq:Sm-shortdistance}
S_m(x_{1},x_{2})\simeq
\begin{cases}
-\dfrac{m}{16\pi\sqrt{u_{1}u_{2}}}\,+\dfrac{m^2}{8\pi}\,s_{12}+\cdots, & D=3,\\[10pt]
C+\dfrac{m^2}{8\pi^2}\,\log\!\big(m\sqrt{u_1u_2}\big)+\dfrac{m^2}{8\pi^2}\log\!\big(s_{12}\big)+\cdots, & D=4,\\[10pt]
-\dfrac{\Gamma\!\left(\tfrac{D-2}{2}\right)}{8\pi^{D/2}(D-4)}\,\dfrac{m^2}{s_{12}^{D-4}}
   +\cdots, & D>4,
\end{cases}
\end{equation}
where $C$ is a constant and ellipses denote terms that vanish in the limit $s_{12}\to 0$. Away from coincident points, $S_{m}$ is smooth. 

In $D=3$, $S_m$ is continuous but not differentiable in angles; this is a very weak singularity, and preliminary analysis suggests that $\omega_0 \prec \omega_m$ should hold.
In $D=5,$ the singularity is power-law, and excitability seems more subtle; the case $D=4$ is marginal between the two.
Preliminary analysis suggests that excitability does hold in $D=4$, though we leave detailed verification of this claim, and an analysis of the case $D \geq 5,$ to future work.

We note that an analogous conjecture is actually proven in the case where one tries to excite $\omega_0$ out of $\omega_m$ on a constant time slice of the causal diamond, instead of on a null slice.
In this case, it was proven by Eckmann and Frohlich that one has $\omega_{0}\vert_{\Sigma}\prec \omega_{m}\vert_{\Sigma}$ for $D\leq 4$ \cite{Eckmann:1974aj} (see also \cite{Conti:Weyl}).
Furthermore, it has been claimed (though we know of no explicit proof) that excitability does not hold on a spatial slice for $D\geq 5$ \cite{Buchholz:1997BV,Conti:Weyl}.

\subsection{Curved backgrounds and free gauge fields}
\label{subsec:QFTCSnulllocal}

The utility of slice-local modular flow was originally envisioned in \cite{Jensen:general-subregions} for applications to semiclassical gravity.
For this setting, it is important to understand how the ideas of the present paper generalize both to curved backgrounds, and to electromagnetic and gravitational fields.

Gauge fields, even if massless, are not necessarily conformally invariant, and do not have vacua analogous to the Klein-Gordon expression used in section \ref{sec:masslessvacuum}.
Nevertheless, in general free gauge theories it is possible to construct states with null-local modular flow analogous to that of the massless scalar vacuum; we perform this construction explicitly in appendix \ref{app:nulllocalemgrav}.
The key point is that except in conformally invariant settings (like the Maxwell field in $D=4$), the state ``$\omega_0$'' with null-local modular flow will not be the vacuum state.
As already emphasized earlier in this section, additional work must be done to determine whether these states can be realized in the vacuum sector.

In general curved spacetimes, the situation is even more subtle due to the absence of a global vacuum state.
Nevertheless, for sufficiently small causal diamonds, the past null boundary $\mathcal{C}^-$ will contain no caustics, and will have the same basic structure as in Minkowski spacetime.
In \cite{Hollands:thesis}, Hollands explained how to generalize the massless vacuum two-point function from equation \eqref{eq:massless-vacuum} to define, on such diamonds, a physically interesting state of the conformally coupled scalar field.
The resulting two-point function still has a singularity of the form $(u_1 - u_2 - i \epsilon)^{-2}$ with respect to the affine parameter along $\mathcal{C}^-,$ but with additional geometric factors that reflect features of the ambient geometry (see also \cite{Dappiaggi:2010iq}).
As already observed in \cite{Hollands:thesis}, this singular structure guarantees that the state has the KMS property with respect to a diffeomorphism of $\mathcal{C}^-$ --- in other words, it has null-local modular flow.
There is no essential difficulty in generalizing this construction to a non-conformally coupled scalar or to free gauge fields, as in appendix \ref{app:nulllocalemgrav}. 
The remaining question is whether this state can be realized in ``physical'' Hilbert space, i.e., in the GNS space corresponding to a global Hadamard state.
As in the case of massless-massive excitations in flat spacetime, we leave the analysis of this question to future work. 

\acknowledgments{We acknowledge stimulating conversations with Jackie Caminiti, Federico Capeccia, Tom Faulkner, Stefan Hollands, Nima Lashkari, Roberto Longo, Gerardo Morsella, and Bob Wald.
This work was supported by the Princeton Gravity Initiative at Princeton University.}

\appendix

\section{Solution smearing in free field theory}
\label{app:solution-smearing}

\begin{figure}
	\centering
	\includegraphics{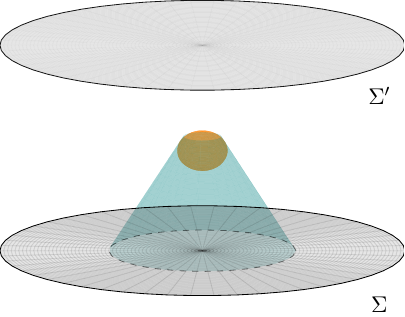}
	\caption{The support of a test function $f$ is shown in orange, and the support of its advanced solution $Af$ is shown in teal. If we surround the support of $f$ with a past Cauchy slice $\Sigma$ and a future Cauchy slice $\Sigma',$ then the boundary terms from equation \eqref{eq:app-vol-integral} are only nonvanishing on $\Sigma.$}
	\label{fig:boundary-term}
\end{figure}

In a globally hyperbolic spacetime $\M$, for any test function $f$, there always exist advanced and retarded solutions to the Klein-Gordon equation with source $f,$ which are uniquely specified by their supports \cite[theorem 6.3.1]{friedlander1975wave}.
The advanced solution $A f$ is supported entirely in the past of the support of $f$; the retarded solution $R f$ is supported entirely in the future of the support of $f.$
The support of an advanced solution is sketched in figure \ref{fig:boundary-term}.
The combination $E f = Af - Rf$ is a solution to the source-free Klein-Gordon equation.

Let $\phi$ be a solution to the Klein-Gordon equation.
We aim to show how the spacetime integral
\begin{equation}
	\phi(f) \equiv \int d^{D} x\, \sqrt{-g}\, \phi f
\end{equation}
can be written in terms of $Ef$ on any Cauchy slice $\Sigma.$
First, we write $f = (\Box - m^2) A f.$
Integrating by parts using the compact support of $f,$ and using the identity $(\Box - m^2) \phi = 0$, we find
\begin{equation} \label{eq:app-vol-integral}
	\int d^{D} x\, \sqrt{-g}\, \phi f
		= \int d^{D} x\, \sqrt{-g} \Del_a (\phi \Del^a Af - Af \Del^a \phi).
\end{equation}
Using Stokes's theorem, this volume integral of a total divergence can be transformed into a surface integral.
We take $\Sigma$ to be in the past of the support of $f$.
Because of the support properties of $Af,$ the boundary term in the integral is localized entirely on $\Sigma$ --- see figure \ref{fig:boundary-term}.
This gives
\begin{equation} \label{eq:appendix-past-surface}
	\int d^{D} x\, \sqrt{-g}\, \phi f
	= \int_{\Sigma} d^{D-1} x\, \sqrt{h} (\phi\, n^{a} \Del_{a} Af - Af\, n^{a}\Del_a \phi).
\end{equation}
On $\Sigma,$ we have $Af = Ef,$ so we may equally well write
\begin{equation} \label{eq:appendix-any-surface}
	\int d^{D} x\, \sqrt{-g}\, \phi f
	= \int_{\Sigma} d^{D-1} x\, \sqrt{h} (\phi\, n^{a} \Del_{a} Ef - Ef\, n^{a}\Del_a \phi).
\end{equation}
While formula \eqref{eq:appendix-past-surface} was only valid for $\Sigma$ in the past of the support of $f,$ equation \eqref{eq:appendix-any-surface} is valid on any Cauchy slice $\Sigma.$
This is because the integral in equation \eqref{eq:appendix-any-surface} is independent of $\Sigma,$ because the divergence
\begin{equation}
	\Del_a (\phi\, \Del^a E f - E f\, \Del^a \phi)
\end{equation}
vanishes identically when $\phi$ and $Ef$ are both solutions to the equations of motion.

\section{Null-local modular flow for Maxwell fields and gravitons}
\label{app:nulllocalemgrav}

In the main text we considered, for simplicity, the massless Klein-Gordon field.
In this theory, the global vacuum state has null-local modular flow when restricted to a causal diamond, and is the unique state in the vacuum sector that has this property.
The null-local modular flow of the state could be seen explicitly from the form of the two-point function in equation \eqref{eq:massless-vacuum}.

In this appendix, we explain how to construct an analogous two-point function for free Maxwell fields and for free gravitons.
As emphasized in section \ref{sec:discussion}, this state will not generally be obtained by a restriction of the global vacuum.

\subsection{Maxwell fields}
\label{app:nulllocalem}

The quantization of the free Maxwell field is very similar to that of the massless Klein-Gordon field.
The essential new feature is gauge invariance.
The dynamical field is the vector potential $A_{a}$.
We will restrict our attention to Minkowski spacetime.

The equation of motion satisfied by the vector potential is 
\begin{equation}
    \Del_a \Del^a A^b - \nabla_{a}\Del^b A^a  = 0.
\end{equation}
Configurations of $A_a$ are identified up to gauge transformations
\begin{equation}
    A_{a}\to A_{a} + \nabla_{a}\chi, \qquad \chi \in C^{\infty}_0(\mathcal{M}).
\end{equation}
Physical observables should respect this invariance, which means that when smearing $A$ against a vector-valued function to obtain an observable, we should smear in such a way that the observable is left unchanged by gauge transformations.
This is accomplished by smearing only against divergence-free vector-valued functions --- one defines the observables
\begin{equation}
    A_{a}[f]
        = \int d^{D} x\, A_a(x) f^a(x) \qquad \Del_a f^a = 0,
\end{equation}
where $f^a$ is compactly supported.

The $*$-algebra of the quantum theory, $\A_{\mathrm{EM}}$, is then defined by taking formal polynomials in the smeared fields $A[f]$, factored by relations analogous to those of section \ref{sec:KG-review}.
In particular, $A(f)$ satisfies the commutator 
\begin{equation}
[A[f_{1}],A[f_{2}]] = -i E[f_{1},f_{2}]
\end{equation}
where $E[f_{1},f_{2}] $ is the advanced-minus-retarded propagator.\footnote{We note that since $A_{a}$ is not gauge invariant, neither is $E$. However, $E[f_{1},f_{2}]$ is gauge invariant for divergence-free $f_{1}^{a}$ and $f_{2}^{b}$.}
The algebra $\mathcal{A}_{\textrm{EM}}$ is expressed in terms of initial data on any Cauchy surface via the conserved symplectic form 
\begin{equation}
\label{eq:sympformEM}
\Omega_{\textrm{EM}}[A_{1},A_{2}] = -\int_{\Sigma}\sqrt{h}d^{D-1}x~n^{a}(A_{1}^{b}F_{2,ab} - A_{2}^{b}F_{1,ab})
\end{equation}
where $F = d A$ is the field strength.
The relation between bulk observables and initial data is given by
\begin{equation}
    A[f] = \Omega_{\textrm{EM}}[A,Ef].
\end{equation}

We now consider the algebra $\mathcal{A}^{\textrm{EM}}(\mathcal{D})$ associated to a causal diamond $\mathcal{D}$ in terms of its initial data on the past null boundary $\mathcal{C}^{-}$.
To describe the free data specifiable on a null surface, it is convenient to choose a preferred gauge in any neighborhood of $\mathcal{C}^{-}$. To describe this gauge we define an advanced time $v$ so that $\ell^{a}=(\partial/\partial v)^{a}$ is a transverse null vector to $\mathcal{C}^{-}$ ($\ell^{a}n_{a}=-1$ and $\ell^{a}\ell_{a}=0$). In the coordinates $(u,v,x^{\perp})$, we choose the gauge
\begin{equation}
\label{eq:gaugecond1}
A_{a}\ell^{a} =A_{v}=0
\end{equation}
To achieve this gauge, the corresponding gauge transformation must satisfy the equation $\ell^{a}\partial_{a}\chi = -A_{v}$; this can be solved as an ODE along each generator $\ell^{a}$ of the null congruence transverse to $\mathcal{C}^{-}$. The solution is unique up to a free function $\chi_{0}(u,x^{\perp})$ on $\mathcal{C}^{-}$, and we may use the remaining gauge freedom to set 
\begin{equation}
A_{a}n^{a}=A_{u}~\hat{=}~ 0 
\end{equation}
where ``$\hat{=}$'' denotes equality after restricting to $\mathcal{C}^{-}$. This is again achieved by solving $n^{a}\partial_{a}\chi_{0}~\hat{=}~-A_{n}$, and the solution is unique up to an angular function $\chi_{0}(x^{\perp})$. In this gauge, the free data on $\mathcal{C}^{-}$ are the angular components of the vector potential $A_{A}(u,x^{\perp})$, up to the residual gauge transformation $A_{A}(u,x^{\perp})\to A_{A}(u,x^{\perp})+\mathscr{D}_{A}\chi_{0}(x^{\perp})$.
The symplectic form \eqref{eq:sympformEM} on $\mathcal{C}^{-}$ becomes
\begin{equation}
\Omega_{\textrm{EM}}[A_{1},A_{2}] = -\int du\, d\Omega~u^{D-4}q^{AB}(A_{1,A}\partial_{u}A_{2,B} - A_{2,A}\partial_{u}A_{1,B}),
\end{equation}
where $q_{AB}$ is the round metric on the unit $(D-2)$-sphere. As in the scalar case, it is convenient to define the following ``rescaled'' field\footnote{\label{footnote:regAtip}In $D=3$ it appears that $\mathcal{A}_{A}$ may blow up at the tip $u=0$ since $u^{\tfrac{D-4}{2}}$ diverges there. However, since the vector potential $A_{a}$ is regular at the tip, it follows that the orthonormal frame component $A_{\hat{A}}=(e_{\hat{A}})^{a}A_{a}$ is also regular at the tip. The frame vector $(e_A)^{a}$ is given by $u^{-1}(\partial/\partial x^{\perp})^{a}$, so $A_{A}$ must be $O(u)$ at the tip.} has a smooth limit to the tip. 
\begin{equation}
\mathcal{A}_{A}(u,x^{\perp})\equiv u^{(D-4)/2}A_{A}(u,x^{\perp})
\end{equation}
in which case the symplectic form becomes 
\begin{equation}
\Omega_{\textrm{EM}}(A_{1},A_{2}) =- \int du\, d\Omega\, q^{AB}(\mathcal{A}_{1,A}\partial_{u}\mathcal{A}_{2,B}-\mathcal{A}_{2,A}\partial_{u}\mathcal{A}_{1,B})
\end{equation}
If we define $F_{a}\equiv u^{(D-2)/2}(Ef)_{a}\vert_{\mathcal{C}^{-}}$ to be the rescaled pullback of $(Ef)_{a}$ to $\mathcal{C}^{-}$, then we may compactly write
\begin{equation}
A[f] = -2\int_{\mathcal{C}^{-}}  du\, d\Omega\, ~F^{A}(u,x^{\perp}) \partial_{u}\mathcal{A}_{A}(u,x^{\perp}).
\end{equation}
where $F^{A}=q^{AB}F_{B}$.
With this relation, the electromagnetic analog of the massless Klein-Gordon vacuum is
\begin{equation}
\label{eq:EMvac1}
\omega^{\textrm{EM}}_{0}(A[f_{1}]A[f_{2}]) = - \frac{1}{\pi} \lim_{\epsilon \to 0^+} \int du_{1}\, du_2\, d\Omega\, q_{AB} \frac{F^{A}_{1}(u_{1},x^{\perp})F^{B}_{2}(u_{2},x^{\perp})}{(u_1 - u_2 - i \epsilon)^2}.
\end{equation}
Following the arguments of section \ref{sec:masslessvacuum}, it is straightforward to verify that this defines a state on $\mathcal{A}^{\textrm{EM}}(\mathcal{D})$ with null-local modular flow on $\mathcal{C}^{-}$.
Furthermore, the proof of theorem \ref{thm:uniquenesstheorem} can be directly generalized to prove that $\omega_{0}^{\textrm{EM}}$ is the unique state (or weight) in its folium with geometric modular flow on $\mathcal{C}^{-}$. 

\subsection{Free gravitons}
\label{app:nulllocalgrav}

We now consider a construction of a state analogous to \eqref{eq:massless-vacuum} for the case of linearized graviton fields.
In many ways, the construction mirrors that of the previous subsection and of section \ref{sec:constructing-states}, so we will focus primarily on the key differences that arise in the quantization of metric perturbations. The key conceptual difference is diffeomorphism invariance, which enters into the discussion of observables in $\mathcal{A}_{\textrm{GR}}$ as well as into the definition of the subregion $\mathcal{D}$ used to defined $\mathcal{A}_{\textrm{GR}}(\mathcal{D})$.

The dynamical field is now a metric perturbation $\gamma_{ab}$ on the the background of Minkowski spacetime, and the equation of motion is 
\begin{equation}
-\frac{1}{2}\nabla_{c}\nabla^{c} \gamma_{ab}
-\frac{1}{2}\nabla_{a}\nabla_{b}\gamma^{c}{}_{c}
+\frac{1}{2}\nabla_{c}\nabla_{a}\gamma^{c}{}_{b}
+\frac{1}{2}\nabla_{c}\nabla_{b}\gamma^{c}{}_{a}=0.
\end{equation}
Two solutions are considered equivalent if they differ by a linearized gauge transformation 
\begin{equation}
\gamma_{ab} \to \gamma_{ab}+\nabla_{(a}\xi_{b)}.
\end{equation}
One can obtain a gauge-invariant observables by smearing against symmetric, divergence-free tensors:
\begin{equation}
\gamma[f] = \int d^{D}x\, \gamma_{ab}(x)f^{ab}(x), \qquad f^{[ab]} = 0 = \Del_a f^{ab}.
\end{equation}
These observables satisfy the covariant commutation relations 
\begin{equation}
[\gamma(f_{1}),\gamma(f_{2})] = -i E[f_{1},f_{2}].
\end{equation}
The algebra $\mathcal{A}_{\textrm{GR}}$ is the algebra of polynomials in $\gamma[f]$, factored by relations analogous to those in section \ref{sec:KG-review}.
The conserved symplectic form is given by 
\begin{equation}
\Omega_{\textrm{GR}}[\gamma_{1},\gamma_{2}]=\frac{1}{16\pi G_{\textrm{N}}}\int_{\Sigma}\sqrt{h}d^{D-1}x~n_{a}P^{abcdef}[\gamma_{2,bc}\nabla_{d}\gamma_{1,ef}-\gamma_{1,bc}\nabla_{d}\gamma_{2,ef}],
\end{equation}
where 
\begin{equation}
P^{abcdef} \equiv \eta^{ae}\eta^{fb}\eta^{cd}
-\tfrac{1}{2}\eta^{ad}\eta^{be}\eta^{fc}
-\tfrac{1}{2}\eta^{ab}\eta^{cd}\eta^{ef}
-\tfrac{1}{2}\eta^{bc}\eta^{ae}\eta^{fd}
+\tfrac{1}{2}\eta^{bc}\eta^{ad}\eta^{ef}.
\end{equation}

The above construction defines an algebra of observables for the graviton field on all of $\mathbb{R}^{D}$. However, we are ultimately interested in the algebra restricted to a subregion $\mathcal{D}\subset \mathbb{R}^{D}$ of the spacetime. In quantum field theory without gravity, we could choose this region arbitrarily. With gravity, the region must be defined in a diffeomorphism-invariant way.
Given such a definition, the subregion $\mathcal{D}$ can be identified both in the Minkowski background and in the perturbed spacetime with metric $\eta_{ab} + \gamma_{ab}.$
Under these conditions, one can assign an algebra $\mathcal{A}_{\textrm{GR}}(\mathcal{D})$ to the subregion.
Solving this problem for general spacetime subregions is the subject of numerous investigations (see, e.g., \cite{Donnelly:2015hta,Donnelly:2016auv,Goeller:2022rsx} and references therein) and is outside the scope of this work. In the following we will simply make the assumption that the region $\mathcal{D}$ is invariantly defined in such a way that (1) it is causal diamond in Minkowski spacetime and (2)  it remains a causal diamond with smooth null boundaries in the perturbed spacetime.

Under these assumptions, we now consider the symplectic form evaluated on $\mathcal{C}^{-}$.
As in the previous subsection, it will be convenient to choose a gauge. The ``double null'' coordinates $(u,v,x^{\perp})$ in Minkowski spacetime are a special case of ``Gaussian null coordinates'' for $\mathcal{C}^{-}$. These coordinates can be defined in the neighborhood of any smooth, null hypersurface (see, e.g., \cite{Hollands:stability} for more details).
Without loss of generality, we choose the perturbed metric to preserve the Gaussian null gauge.
In the coordinates $(u,v,x^{\perp})$, this implies 
\begin{align}
&\gamma_{ab}\ell^{a}\ell^{b}=\gamma_{ab}\ell^{a}(e)^{b}_{A}=\gamma_{ab}\ell^{a}n^{b}=0  \nonumber \\
&\gamma_{ab}n^{a}n^{b}~\hat{=}~\gamma_{ab}n^{a}(e)^{b}_{A}~\hat{=}~\ell^{a}\nabla_{a}(\gamma_{bc}n^{b}n^{c})=0.
\end{align}
In these equations, we have $\ell^{a}=(\partial/\partial v)^{a}$ and $n^{a}=(\partial/\partial u)^{a}$, and $(e)^{a}_{A}$ with ($A=1,\dots,D-2$) is the frame tangent to constant-$u$ cross-sections of $\mathcal{C}^{-}$ and unit normalized with respect to the round metric $q_{AB}$ on the unit sphere.

In this gauge, the only non-trivial components of $\gamma_{ab}$ on $\mathcal{C}^{-}$ are the angular components $\gamma_{AB}(u,x^{\perp})$. The trace $ q^{AB}\gamma_{AB}$ is controlled by the perturbed expansion
$\delta \theta = \frac{1}{2}\partial_{u}(u^{-2}q^{AB}\gamma_{AB})$ which satisfies Raychaudhuri’s equation \begin{equation}
\partial_{u}\delta \theta + \frac{2}{u}\delta \theta = 0 
\end{equation}
The general solution is $\delta \theta = C(x^{\perp})/u^{2}$. Since $\gamma_{ab}(x)$ is smooth in a neighborhood of the tip of $\mathcal{C}^-$, the perturbed expansion cannot diverge faster than the background expansion $\theta=(D-2)/u $ of $\mathcal{C}^{-}$ in the limit $u\to 0$. Therefore, $C(x^{\perp})=0$ and so $\partial_{u}(u^{-2}q^{AB}\gamma_{AB})=0$. Using the fact that $q^{AB}\gamma_{AB}$ vanishes at the edge of $\mathcal{C}^-$, it follows that $q^{AB}\gamma_{AB}=0$ on $\mathcal{C}^{-}$. The free data on $\mathcal{C}^{-}$ consists of the trace-free part of $\gamma_{AB}$ with respect to $q_{AB}$ 
\begin{equation}
s_{AB}(u,x^{\perp})\equiv \bigg(q_{A}{}^{C}q_{B}{}^{D}-\frac{1}{D-2}q_{AB}q^{CD}\bigg)\gamma_{CD}(u,x^{\perp})
\end{equation}
The symplectic form, in terms of this data, is
\begin{equation}
\Omega_{\textrm{GR}}[\gamma_{1},\gamma_{2}] = \frac{1}{32\pi G_{\textrm{N}}}\int_{\mathcal{C}^{-}}du d\Omega~u^{D-6}q^{AC}q^{BD}(s_{1,AB}\partial_{u}s_{2,CD} - s_{2,AB}\partial_{u}s_{1,CD})
\end{equation}
Defining the rescaled fields\footnote{Since $s_{\hat{A}\hat{B}}=s_{ab}(e_{\hat{A}})^{a}(e_{\hat{B}})^{b}$ is regular at the tip, we have $s_{AB}\sim O(u^{2})$ near $u=0$ (see footnote \ref{footnote:regAtip}).} 
\begin{equation}
\mathcal{S}_{AB} \equiv u^{\frac{D-6}{2}}s_{AB} 
\end{equation}
where 
\begin{equation}
\gamma[f] = -\frac{1}{16\pi G_{\textrm{N}}} \int  du\, d\Omega\, ~F^{AB}(u,x^{\perp}) \partial_{u}\mathcal{S}_{AB}(u,x^{\perp}).
\end{equation}
where $F_{ab}=u^{\frac{D-6}{2}}(Ef)_{ab}\vert_{\mathcal{C}^{-}}$ in Gaussian null gauge we obtain the gravitational analog of the Gaussian state $\omega_{0}$ is the zero-mean Gaussian state with two-point function given by 
\begin{align}
\begin{split}
\label{eq:GRvac1}
\omega^{\textrm{GR}}_{0}(\gamma[f_{1}]\gamma[f_{2}]) = \frac{1}{32\pi^{2} G_{\textrm{N}}}\int du_{1}\,d\Omega_1\, du_{2}\,d\Omega_2~& F_{1,AB}(u_{1},x_{1}^{\perp})F^{AB}_{2}(u_{2},x_{2}^{\perp})~ \\
& \times \frac{\delta_{\mathbb{S}^{D-2}}(x_{1}^{\perp},x_{2}^{\perp})}{(u_{1}-u_{2}-i0^{+})^{2}}.
\end{split}
\end{align}

It is straightforward to show that $\omega_{0}^{\textrm{GR}}$ defines a state on $\mathcal{A}^{\textrm{GR}}(\mathcal{D})$ which has geometric modular flow on $\mathcal{C}^{-}$. Since the linearized gravitational field is not conformally invariant, $\omega_{0}^{\textrm{GR}}$ is not the restriction of the vacuum state to $\mathcal{D}$.

\bibliographystyle{JHEP}
\bibliography{bibliography}

\end{document}